\documentclass[11pt]{article}
\usepackage{amssymb}
\usepackage{amsmath}
\usepackage[all]{xy}
\usepackage[dvips]{graphicx}
\numberwithin{equation}{section}

\def\be{\begin{equation}}
\def\ee{\end{equation}}
\def\ba{\begin{align}}
\def\ea{\end{align}}
\def\beq{\begin{eqnarray}}
\def\eeq{\end{eqnarray}}
\def\a{\alpha}
\def\b{\beta}
\def\p{\partial}
\def\pbar{\bar \partial}
\def\xbar{\bar x}
\def\ybar{\bar y}
\def\wbar{\bar w}
\def\zbar{\bar z}

 \input{epsf}

 \usepackage{epsfig}

\begin{document}

\title{\Large{\bf Asymptotic Symmetries of String Theory on $AdS_3 \times S^3$  
\\ with Ramond-Ramond Fluxes
}} 
\author{Sujay K. Ashok$^{a,b}$, Raphael Benichou$^{c}$ and Jan Troost$^{c}$ }
\date{}
\maketitle
\begin{center}
  $^{a}$Institute of Mathematical Sciences\\
  C.I.T Campus, Taramani\\
  Chennai, India 600113\\
  \vspace{.3cm}
  $^{b}$Perimeter Institute for Theoretical Physics\\
  Waterloo, Ontario, ON N$2$L$2$Y$5$, Canada \\
  \vspace{.3cm}
  $^{c}$Laboratoire de Physique Th\'eorique \\
Unit\'e Mixte du CRNS et
    de l'\'Ecole Normale Sup\'erieure \\ associ\'ee \`a l'Universit\'e Pierre et
    Marie Curie 6 \\ UMR
    8549  \footnote{Preprint LPTENS-09/19.} \\ \'Ecole Normale Sup\'erieure \\
  $24$ Rue Lhomond Paris $75005$, France
\end{center}

 \begin{abstract}
   String theory on $AdS_3$ space-times with boundary conditions that
   allow for black hole states has global asymptotic symmetries which
   include an infinite dimensional conformal algebra. Using the
   conformal current algebra for sigma-models on $PSU(1,1|2)$,
 we explicitly construct the R-symmetry and
   Virasoro charges in the worldsheet theory describing string theory
   on $AdS_3 \times S^3$ with Ramond-Ramond fluxes. We also indicate
   how to construct the full boundary superconformal algebra. The
   boundary superconformal algebra plays an important role in
   classifying the full spectrum of string theory on $AdS_3$ with
   Ramond-Ramond fluxes, and in the microscopic entropy counting in
   D1-D5 systems.
\end{abstract}

\newpage

\tableofcontents

\section{Introduction}
The holographic correspondence between gauge theories and gravitational
theories has shed light both on non-perturbative quantum gravity and
on strong coupling phenomena in gauge theory.  The quintessential
example of a pair of theories related by holography is the four-dimensional 
${\cal N}=4$ super Yang-Mills theory, and type IIB string theory on
the  $AdS_5 \times S^5$ space-time with
Ramond-Ramond flux
\cite{Maldacena:1997re}.  The further development of the
correspondence has been somewhat hampered by the difficulty in solving
string theory on non-trivially curved backgrounds with Ramond-Ramond
flux (except in a plane wave limit
\cite{Berenstein:2002jq, Metsaev:2002re}). A tool that
has stimulated significant progress is the integrability of the spectrum of the
dilatation operator in ${\cal N}=4$ super Yang-Mills theory
\cite{Minahan:2002ve} as well as the
integrability of the bulk worldsheet
$\sigma$-model \cite{Bena:2003wd, Gromov:2009tv}. 
The integrable structure is coded in the
existence of an infinite number of non-local charges.

In a particular instance of the holographic correspondence, one can
make headway via the existence of an infinite number of {\em local}
charges. Indeed $AdS_3 \times S^3$ bulk superstring theory is dual to
a conformal field theory in two dimensions. It therefore has the
exceptional property of having an infinite number of local charges
which extend the finite dimensional isometry group
\cite{Brown:1986nw}. We believe it is important to construct the
symmetry algebra in the bulk string theory explicitly and to exploit it
maximally in classifying the spectrum.  It will moreover be
interesting to see how the local infinite dimensional symmetry algebra
intertwines with the integrability of the model. 

Quantum gravity on $AdS_3$ space-times supplemented with boundary
conditions that allow for black hole solutions has an
asymptotic symmetry group which includes the two-dimensional conformal
algebra \cite{Brown:1986nw}.  For string theory on an $AdS_3$ background
with Neveu-Schwarz-Neveu-Schwarz flux, the space-time symmetry
generators were explicitly constructed in terms of worldsheet
operators in \cite{Giveon:1998ns,  de Boer:1998pp, Kutasov:1999xu}.

Here we concentrate on $AdS_3 \times S^3$ backgrounds of string theory
with non-zero Ramond-Ramond flux with eight or sixteen
supercharges. These backgrounds arise as near-brane geometries of
 D1-D5 brane
configurations (which may also include fundamental strings
and NS5-branes). After introducing a third charge, 
the conformal symmetry is
central in the microscopic counting that reproduces the Bekenstein-Hawking
area formula for the black hole entropy. We believe it is
 useful to exhibit this
symmetry directly and explicitly in the D1-D5 near-brane geometry.

To that end we work in the hybrid formalism of \cite{BVW} which
renders eight spacetime supercharges manifest. In that formalism the
central part of the worldsheet model is the sigma-model on the
$PSU(1,1|2)$ supergroup (or rather, its universal cover), which
contains a bosonic $AdS_3 \times S^3$ subspace.  The main building
block in constructing the vertex operators of the spacetime symmetry
generators are the curents of the $PSU(1,1|2)$ supergroup model which
satisfy a conformal current algebra found in \cite{ABT}. Technically,
our analysis is a non-chiral version of the construction of the super
Virasoro algebra in \cite{de Boer:1998pp, Kutasov:1999xu,
  Giveon:2001up} for the case of F1-NS5 backgrounds. For interesting
studies of the supergroup sigma-model on $PSU(1,1|2)$ at the
Wess-Zumino-Witten points and beyond in the context of string theory
we refer to
\cite{Bershadsky:1999hk, Gotz:2005ka, Gotz:2006qp, Quella:2007sg}.

This article is organized as follows. In section \ref{model} we review
key features of the worldsheet sigma-model in the hybrid
formalism. The worldsheet current algebra of the model is recalled in
section \ref{worldsheetcurrentalgebra}. The construction of the vertex
operators for the R-current algebra generators in space-time and the
calculation of their operator product expansion (OPE) is performed in
section \ref{Rcurrents}. In section \ref{Virasoro} we indicate how to
construct the full space-time super Virasoro algebra and discuss the
properties of the operator that plays the role of the central extension in the spacetime algebra. We summarize and make our concluding remarks in section \ref{conclusions}. Some technical details regarding the
worldsheet action and certain aspects regarding worldsheet operators
and their OPEs are collected in the appendices.

\section{String Theory on $AdS_3 \times S^3$ with Ramond-Ramond Fluxes}
\label{model}

In this section, we briefly review the hybrid formalism, the supergroup sigma-model that describes the $AdS_3 \times S^3$ background with Ramond-Ramond fluxes, and its relation to the near-brane geometries. We also discuss the BRST operator of the worldsheet theory in the hybrid formalism, whose cohomology determines the physical string Hilbert space.

\subsection{The Hybrid Formalism}

The hybrid formalism introduced in \cite{Berkovits:1994vy} allows one to
covariantly quantize string theory with Ramond-Ramond fluxes in a
six-dimensional space-time.  It renders eight supercharges
in space-time manifest\footnote{There exists another formulation of
  the six-dimensional hybrid string which renders sixteen supercharges
  manifest \cite{Berkovits:1999du}. In this formalism,
  the space-time $AdS_3\times S^3$ is embedded in a super-coset. We concentrate on
 the formalism defined in \cite{Berkovits:1994vy}
  in which the manifold $AdS_3\times S^3$ is embedded in a supergroup
  \cite{BVW}.}.  The formalism is based on defining space-time
fermions $\theta^{a\alpha}$ and their conjugate momenta $p_{a\alpha}$ in terms of
the spin-fields in the RNS formalism. There are six bosons
corresponding to the six space-time directions as well as two
chiral interacting bosons $\rho$ and $\sigma$ (and their right-moving
counterparts $\bar{\rho}$ and $\bar{\sigma}$) related to the bosonized
ghost systems of the RNS formalism.

In the $AdS_3 \times S^3$ background the ghost fields $\phi=-\rho-i
\sigma$ and $\bar \phi = -\bar \rho - i \bar \sigma$ are respectively
chiral and anti-chiral (up to interaction terms).  The ghosts appear in the action as
exponentials of the fields 
$\phi$ and $\bar \phi$ with positive exponent.  In other
words,  the
worldsheet action is a perturbation series in
the variables $e^{\phi}$ and $e^{\tilde{\phi}}$.  In the presence of non-zero
Ramond-Ramond flux, the worldsheet Lagrangian
is quadratic in the fermionic momenta $p_{a\alpha}$. It is then possible
to integrate out the fermionic momenta, so that the action only
depends on the bosonic and fermionic coordinates (as well as the
ghosts) \cite{BVW}.

\subsection{The Supergroup Sigma Model}

We will work to lowest order in the ghost exponentials.  At this
order, the action pertaining to the six-dimensional space $AdS_3
\times S^3$ is a non-linear sigma-model with target space the
(covering of the) supergroup manifold $PSU(1,1|2)$ \cite{BVW}.
The supergroup $PSU(1,1|2)$ has a maximal bosonic subgroup which is
$SU(1,1) \times SU(2)$.  There is a corresponding $psu(1,1|2)$
superalgebra which is a particular real form of the $psl(2|2)$
superalgebra. The latter is defined by its generators and their
(anti-)commutation relations are given by
\begin{align} \label{KKcomm}
[K_{ab},K_{cd}] &= \delta_{ac}K_{bd}-\delta_{ad}K_{bc}-\delta_{bc}K_{ad}+\delta_{bd}K_{ac} \cr
\left[ K_{ab},S_{c \alpha} \right] &= \delta_{ac}S_{b\alpha} - \delta_{bc}S_{a\alpha} \cr
\{S_{a\alpha},S_{b \beta}\} &= \frac{1}{2} \epsilon_{\alpha \beta}\epsilon_{abcd} K^{cd} \,.
\end{align}
where $a,b$ are vector indices of $so(4) \sim sl(2)\times sl(2)$.  The
generators $K_{ab}=-K_{ba}$ are the bosonic generators. There is an
outer automorphism algebra $sl(2)_{out}$ (which has the real form
$su(2)_{out}$ in the case of the algebra $psu(1,1|2)$) and the indices
$\alpha, \beta$ run over the states in a doublet of $sl(2)_{out}$.
The tensor $\epsilon_{\alpha \beta}$ is anti-symmetric.  Under the
bosonic subalgebra and the outer automorphism algebra $sl(2)\oplus
sl(2) \oplus sl(2)_{out}$ the six bosonic generators transform as
$(3,1,1) + (1,3,1)$ and the eight fermionic generators as $(2,2,2)$
for a total of fourteen generators that span the adjoint of
$psl(2|2)$.
 
There is no fundamental representation of the $psl(2|2)$ algebra, but
in the appendix we give an explicit $4 \times 4$ matrix
parameterization of the $sl(2|2)$ algebra (which is the $psl(2|2)$
algebra augmented with a central bosonic generator). Explicit
calculations can be performed with the matrix parameterization of this
algebra, and results for the $psl(2|2)$ model are obtained by dividing
out by the central generator.

The action of the non-linear sigma-model on the supergroup is:
\begin{align}\label{ourmodel}
S &= S_{kin} + S_{WZ}\cr
S_{kin} &=  \frac{1}{ 16 \pi f^2}\int d^2 z Tr'[- \partial^\mu g^{-1}
\partial_\mu g]
\cr
S_{WZ} &= - \frac{ik}{24 \pi} \int_B d^3 y \epsilon^{\alpha \beta \gamma}
Tr' (g^{-1} \partial_\alpha g g^{-1} \partial_\beta g   g^{-1} \partial_\gamma g )
\end{align}
where $g$ takes values in the supergroup $PSU(1,1|2)$ and $Tr'$ indicates the 
non-degenerate bi-invariant metric. It can be thought of as the supertrace in
the $su(1,1|2)$ superalgebra. We parameterize the group element
$g \in SU(1,1|2)$ as 
\be 
g = e^{\alpha} e^{\theta^{a\alpha}S_{a\alpha}} g_{S^3} g_{AdS_3}\,,
\ee
where the first factor represents the $U(1)$ to be divided out, the
second factor the fermions, the third one an element of the group
$SU(2)$ and the last factor an element of the group
$SL(2,\mathbb{R})$. We spell out the non-linear supergroup sigma-model
action in much more detail in terms of a global coordinate system in
appendix \ref{componentaction}. From the kinetic term in equation
\eqref{fulllagrange}, we find that the quadratic fermionic term in the
action takes the form
\be
S_{fermionic} = \frac{1}{4 \pi f^2} \int d^2 z\ \delta_{ab}\epsilon_{\a\b} 
\partial\theta^{a\a}\bar\partial \theta^{b\b} + \ldots \,
\ee
The main advantage of the hybrid formalism 
compared to the Green-Schwarz superstring
is that it is covariantly
quantizable.

\subsection{The Brane Configuration}

The coefficient $\frac{1}{f^2}$ multiplying the kinetic term is the
square of the spacetime radius.  The coefficient $k$ multiplying the
Wess-Zumino term is quantized. It is equal to the number of units of
Neveu-Schwarz-Neveu-Schwarz flux on the three-sphere.  When $AdS_3
\times S^3$ is realized as the near-brane geometry of a system of
NS5-F1-D5-D1 branes, the couplings $\frac{1}{f^2}$ and $k$ are related
to the number of NS5-branes $Q_{NS5}$ and D5-branes $Q_{D5}$ through
the formulas \cite{BVW}:
\begin{eqnarray}
\frac{1}{f^2} &=& \sqrt{Q_{NS5}^2 + g_s^2 Q_{D5}^2}\cr
k &=& Q_{NS5}\,, 
\end{eqnarray}
where $g_s$ is the ten-dimensional string coupling constant. Note that
since the D5-branes are a factor of $1/g_s$ lighter than the
NS5-branes, they curve the geometry less strongly. Supersymmetry
requires the relative number of D1-branes compared to F1-strings to be
equal to the relative number of D5-branes compared to NS5-branes.  We
thus have:
\be
(Q_{NS5},Q_{D5}) = Q_5 (p,q)  \quad \text{and}\quad (Q_{F1},Q_{D1})=Q_1 (p,q)\,.
\ee
and via the attractor mechanism $Q_1$ (divided by $Q_5$) fixes the
volume of the compactification manifold. For a type IIB superstring
background with sixteen or eight supercharges, the internal manifold
can be either
a four-torus or a K3 manifold.

\subsection{The Physical Hilbert Space}

The worldsheet theory also contains an $N=4$ superconformal
 model
at central charge $c=6$ associated to the compactification manifold.
Via the ghosts, the principal chiral model with Wess-Zumino term is
coupled to 
the compact theory. The correlation functions of the model can be
defined in terms of the prescription for computing $N=4$ topological
string amplitudes \cite{Berkovits:1994vy}. 

 The physical string Hilbert space is given in
terms of a set of constraints that can be defined in terms of an $N=2$
superconformal algebra. In particular, physical states belong to the
cohomology of the charges associated to the BRST currents $G^+$ and
$\bar G^+$. At zeroth order in the ghost exponential $e^\phi$, the
holomorphic BRST current $G^+$ reads: \be\label{BRSTcurrent} G^+ =
e^{i \sigma} \left( T_{PSU(1,1|2)} + \frac{1}{2}(\partial
  \phi \partial \phi + \partial^2 \phi) \right) + G^+_C +
\mathcal{O}(e^\phi) \ee where $G^+_C$ depends only on the
compactification variables and the ghosts, and $T_{PSU(1,1|2)}$ is the
holomorphic stress-energy tensor of the supergroup sigma-model, which is given
in terms of a generalized Sugawara construction \cite{ABT}.

Our goal is to construct, in the worldsheet theory, the vertex
operators for the generators of the asymptotic symmetry group for the
$AdS_3 \times S^3$ string theory.
We will show that these vertex operators are BRST closed with respect
to the charge associated to the current \eqref{BRSTcurrent}. As such,
we will be able to insert these vertex operators in correlation
functions to generate space-time Ward-identities.

\section{The Worldsheet Conformal Current Algebra}
\label{worldsheetcurrentalgebra}
In the previous section we reviewed that the supergroup sigma-model on 
$PSU(1,1|2)$ is the central building block for $AdS_3 \times S^3$ string
theory with Ramond-Ramond flux in the hybrid formalism. The supergroup
sigma-model has zero Killing form and it therefore falls into the class
of models for which the worldsheet conformal
 current algebra was analyzed in \cite{ABT}. Since the worldsheet
current algebra will be the central technical tool in constructing the
space-time Virasoro algebra, we review it here. We also introduce
current algebra primary fields and some useful notations.

\subsection{Conformal Current Algebra}
From the action \eqref{ourmodel} we can calculate the classical
currents associated to the invariance of the theory under left
multiplication of the field $g$ by a group element in $G_L$ and right
multiplication by a group element in $G_R$.
The classical $G_L$ currents are given by
\begin{align}\label{normeqn}
j_{L,z} &= c_+ \partial g g^{-1}\cr
j_{L,\bar{z}} &= c_- \bar{\partial} g g^{-1} \,,
\end{align}
where the constant $c_+$ and $c_-$ are given in terms of the couplings
by: \be\label{c+-} c_{\pm} = -\frac{(1\pm kf^2)}{2f^2} \,.  \ee
Similarly, we also have the left-invariant currents that generate
right multiplication:
\begin{align}
j_{R,z} &= -c_- g^{-1} \p g \cr
j_{R,\bar z} &= -c_+ g^{-1} \bar \p g\, .
\end{align}
The operator product expansions satisfied by the left currents have been
derived in \cite{ABT}, where now, $a$ denotes a super Lie algebra 
valued index:
\begin{align}\label{euclidOPEs}
j_{L,z}^a (z) j_{L,z}^b (0) & \sim  \kappa^{ab} \frac{c_1}{z^2} + {f^{ab}}_c \left[ \frac{c_2}{z} j_{L,z}^c(0)+ (c_2-g) \frac{\bar{z}}{z^2} j_{L,\bar{z}}^c(0) \right] + \ldots \cr
j_{L,\bar{z}}^a (z) j_{L,\bar{z}}^b (0) & \sim  \kappa^{ab}  \frac{c_3}{\bar{z}^2} 
  + {f^{ab}}_c \left[  \frac{c_4}{\bar{z}} j_{L,\bar{z}}^c(0) + (c_4-g)\frac{z}{\bar{z}^2} j_{L,z}^c(0)\right] +\ldots \cr
j_{L,z}^a (z) j_{L,\bar{z}}^b(0) & \sim \tilde{c}\kappa^{ab} 2\pi \delta^{(2)}(z-w) + {f^{ab}}_c  \left[  \frac{(c_4-g)}{\bar{z}} j_{L,z}^c(0)
+\frac{(c_2-g) }{z} j_{L,\bar{z}}^c(0) \right] + \ldots
\end{align}
The ellipses refer to subleading terms proportional to the derivatives
of the current. We will only need the leading singular behaviour of
the operator product expansions to derive the spacetime superconformal
algebra.  The right current components $j_{R,z}$ and $j_{R,\zbar}$
satisfy similar operator product expansions with the holomorphic
coordinates replaced by anti-holomorphic ones. In appendix
\ref{componentaction} we give explicit expressions for left-invariant
right-moving currents in the $PSU(1,1|2)$ supergroup model in a
particular coordinate system.  For the supergroup non-linear
sigma-model in equation \eqref{ourmodel}, the coefficients of the
conformal current algebra, expressed purely in terms of $c_{\pm}$, are
given by \cite{ABT}
\begin{align}\label{candg}
c_1 &= -\frac{c_+^2}{c_++c_-} \qquad\qquad\qquad c_3 = -\frac{c_-^2}{c_++c_-} \cr
c_2 &= i \frac{c_+(c_++2c_-)}{(c_++c_-)^2} \qquad\qquad c_4 =  i \frac{c_-(2c_++c_-)}{(c_++c_-)^2}  \cr 
g &= i \frac{2c_+c_-}{(c_++c_-)^2} \qquad\qquad\qquad \tilde{c} = \frac{ c_+ c_-}{c_++c_-} \,,
\end{align}
where $c_{\pm}$ are the factors defined in \eqref{c+-} that normalize
the currents. For future purposes, we note that due to the existence
of the elementary group valued field $g$ these coefficients satisfy
the equations\footnote{Roughly speaking, we can think of $\log g$ as 
having a logarithmic operator product expansion with itself, as in an abelian
theory. The current component operator product expansions are then
derivatives of this more basic logarithmic OPE. That gives rise to relations
between $c_1,c_3$ and $\tilde{c}$ (which leads to the first two identities,
in a particular normalization for the currents). 
The Ward identity for the left translation of the group
valued field $g$ gives rise to the third identity.}
\begin{align}\label{relationsCs}
\frac{\tilde{c}-c_1}{c_+} = 1 &= \frac{\tilde{c}-c_3}{c_-} \qquad
\text{and}\qquad c_2+c_4-g = i.
\end{align}

\subsection{The $SL(2,\mathbb{R})\times SL(2,\mathbb{R})$ Global Symmetry}

The zero-mode of the global current
\be
{\cal J}_L^a = -i(j_{L,z}^a-j_{L,\zbar}^a) \,
\ee
generates the symmetry which is the left translation of a supergroup element
 by an element of the supergroup.
Restricting the index $a$ to just the $SL(2,\mathbb{R})$ directions, we obtain the 
generators of the  $SL(2,\mathbb{R}) \times SL(2,\mathbb{R})$ isometry group of $AdS_3$,
which is a subgroup of the left/right space-time Virasoro algebra.
In holography, the $AdS_3$ submanifold of the supergroup plays a special role, which makes it
useful to decompose observables in the spacetime theory 
in terms of representations of this subgroup. We introduce
 auxiliary complex variables $(x,\xbar)$ in terms of which
the global $SL(2,\mathbb{R}) \times SL(2,\mathbb{R})$ generators are expressed simply in
terms of differential operators.

A given observable $O(x,\bar{x})$ in an irreducible representation of
the group, can be generated from its value at $x=\xbar=0$ by acting
with the lowering generators of the $sl(2, \mathbb{R})$ algebra ${\cal
  J}_{L,R}^-$
\be
{\cal O}(x,\xbar) = e^{-x{\cal J}_{L,0}^- -\xbar {\cal J}_{R,0}^-}{\cal O}(0,0) \,,
\ee
where the $0$ index indicates the zero mode of the current. In
particular, we can apply this operation to a current component in the
adjoint representation of the group:
\begin{align}\label{defjLz}
j_{L,z}(x;z)&\equiv j_{L,z}^+(x,;z,\zbar) = e^{-x {\cal J}^-_{L,0}} j_{L,z}^+(z,\zbar) e^{x {\cal J}^-_{L,0}} \cr
&= j^+_{L,z}(z,\zbar)-2xj^3_{L,z}(z,\zbar)+x^2j^-_{L,z}(z,\zbar)\,.
\end{align}
Similar relations hold for the $\zbar$ component as well as the
right-currents. We have for instance:
\be
j_{R,\zbar}(\xbar;z) =  j^+_{R,\zbar}(z,\zbar)-2\xbar j^3_{R,\zbar}(z,\zbar)+\xbar^2j^-_{R,z}(z,\zbar)\,.
\ee
In space-time the $+$ component of the currents $j^+_L$
is of conformal weight
$(-1,0)$ with respect to the zero-modes of the Virasoro
algebras.

\subsection{Left Current Algebra Primaries}

It will be useful to define fields which are primaries with respect to
the left current algebra.  A left primary field $\phi$ with respect to
the current algebra \eqref{euclidOPEs} is a field satisfying the operator product expansions:
\begin{align}  
 j_{L,z}^a(z,\bar z) \phi(w,\bar w) &= - \frac{c_+}{c_+ + c_-} t^a \frac{\phi(w,\bar w)}{z-w} + \text{less singular} \cr
 j_{L,\bar z}^a(z,\bar z) \phi(w,\bar w) &=- \frac{c_-}{c_+ + c_-} t^a \frac{\phi(w,\bar w)}{\bar z-\bar w} + \text{less singular} 
\end{align}
where the matrices $t^a$ are the generators of the Lie super-algebra
taken in the representation in which $\phi$ transforms. The
coefficients are fixed by the global transformation properties of the
field and the demand that the field $\phi$ have trivial operator
product expansion with the Maurer-Cartan operator $c_+\partial j_{\bar
  z}^a - c_-\bar \partial j_z^a + \frac{i}{2} {f^a}_{bc} (:j_z^c j_{\bar
  z}^b: + (-1)^{bc} :j_{\bar z}^b j_z^c: )$.

\subsection{Representations of the Global Bosonic Symmetry Group}

For later purposes, we define a bosonic field $\Phi_h$ transforming in
the representation $(\mathcal{D}^+_h,0)$ of the bosonic subgroup
$SL(2,\mathbb{R})\times SU(2)$ of the supergroup\footnote{In our
  conventions, the quadratic casimir of the representation
  $\mathcal{D}^+_h$ of $SL(2,\mathbb{R})$ is $h(h-1)$.}. When we
continue $AdS_3$ to Euclidean signature it will function as the unique
bulk-to-boundary propagator for a scalar field coupling to a
space-time operator of dimension $h$. We think of the bosonic field
$\Phi_h(y, \bar{y};w,\bar{w})$ as parameterizing (via the variables
$y, \bar{y}$) a subspace of a representation of the supergroup
corresponding to a left primary field, in which case the operator
product expansions of this field with the components of the current
$j_{L}(x;z,\bar z)$ defined in equation \eqref{defjLz} read:
\begin{align} \label{jLwithPhi}
j_{L,z}(x;z,\bar z) \Phi_h(y,\bar y; w,\bar w) &= \frac{c_+}{c_+ +c_-} \frac{1}{z-w}[(y-x)^2\partial_y + 2h(y-x)]\Phi_h \cr
j_{L,\bar z}(x;z,\bar z) \Phi_h(y,\bar y; w,\bar w) &= \frac{c_-}{c_+ +c_-} \frac{1}{\bar z-\bar w}[(y-x)^2\partial_y + 2h(y-x)]\Phi_h.
\end{align}
We have used the fact that the generators $t^a$ in the representation
in which $\Phi_h$ transform can be written as the following
differential operators:
\be\label{tDiffOps} t^3 =
-x \partial_x -h \quad ; \quad t^+ = -x^2 \partial_x - 2hx \quad ; \quad
t^- = -\partial_x \,, \ee
in the conventions of \cite{Kutasov:1999xu}.
It is also useful to rewrite the operator product expansion of the
$SL(2,\mathbb{R})$ components of the left-currents as follows:
\begin{multline}
j_{L,z}(x;z)\cdot j_{L,z}(y;w) = \frac{c_1(x-y)^2}{(z-w)^2} +\frac{c_2\, {f^{ab}}_c}{i(z-w)}\left[(y-x)^2\p_y - 2(y-x)\right] j_{L,z} \cr
+ \frac{(c_2-g)\, (\zbar-\wbar){f^{ab}}_c}{i(z-w)^2}\left[(y-x)^2\p_y - 2(y-x)\right] j_{L,\zbar}+...
\end{multline}
After these worldsheet preliminaries, we turn to the construction of the space-time R-current.

\section{The Spacetime R-Current}\label{Rcurrents}

\subsection{R-Symmetry Generators from Non-trivial Diffeomorphisms}

We consider the space-time background $AdS_3 \times S^3 \times X$ which is a
solution of type IIB string theory. The compact space $X$ can be
either $T^4$ or $K3$.  In the following discussion we will focus on
the $AdS_3 \times S^3$ factor.  The massless excitations of the string
give the supergravity multiplet and one tensor multiplet of $D=6$,
$N=2$ supergravity.  In the hybrid formalism, this has been shown in
detail in \cite{BVW, Dolan:1999dc} by computing the cohomology of the
BRST operator associated to the current in equation
\eqref{BRSTcurrent}.
If we further reduce the theory down to $AdS_3$,
the fluctuations of the metric with one index in $AdS_3$ and one index
in $S^3$ give rise to an $SU(2)$-valued massless vector field in three
dimensions. These gauge bosons in the bulk are associated with
dimension one currents on the boundary.  The space-time conformal
field theory exhibits, in particular, an $SU(2)_L \times SU(2)_R$
R-symmetry group corresponding to the bulk isometries of the
three-sphere.

The R-symmetry algebra extends to a current algebra on the
boundary. From the bulk point of view, this is due to the existence of
diffeomorphisms which act non-trivially on the space of bulk
configurations with particular boundary conditions. Diffeomorphisms
that fall off slowly generate a non-trivial asymptotic symmetry group.
For the supersymmetric backgrounds under consideration, these
diffeomorphisms were analyzed in \cite{de
  Boer:1998ip, Henneaux:1999ib} in the supergravity approximation.

Diffeomorphisms that induce a non-vanishing transformation at infinity
act non-trivially on the Hilbert space of the theory. We parameterize
$AdS_3$ with the Gaussian coordinate system  $(\gamma,\bar
\gamma,\phi)$ on $SL(2,\mathbb{R})$ which admits an analytic continuation to euclidean
$AdS_3$ in Poincar\'e coordinates. In these coordinates the $AdS_3$ metric takes the form:
\be ds^2_{AdS_3} = d\phi^2 + e^{2\phi}d\gamma d \bar \gamma. \ee
The asymptotic symmetry group is generated by diffeomorphisms for
which the generating vector field $\xi$ behaves near the boundary as
\be \xi^{\mu} = f(\gamma,\bar \gamma) + \mathcal{O}(e^{-\phi})\,, \ee
where the index $\mu$, runs over the six indices corresponding to the $AdS_3$ and $S^3$ directions. 
We can further expand the function $f(\gamma,\bar \gamma)$ in powers of $\gamma $ and $\bar \gamma$.  Since the
coordinates $(\gamma,\bar \gamma)$ parameterize radial slices, this is
equivalent to a Fourier mode expansion in the boundary coordinates.
The integrated string vertex operator for an $SU(2)$ gauge boson
is\footnote{Whether we write down the vertex operator in terms of the
  left- or of the right-current is a matter of convention. Indeed
  both one-forms associated to the left- and right-currents generate a
  suitable basis for the cotangent bundle in spacetime.}
\be 
\int d^2z \left(
j_{L,z}^a j_{L,\bar z}^m + j_{L,\bar z}^a j_{L, z}^m 
\right) g_{am} \,,
\ee
where, from now on, we specify $a$ to be the $SU(2)$ index, $m$ is an $SL(2,\mathbb{R})$ index, and $g_{\mu \nu}$ is the six-dimensional metric.  Under a diffeomorphism generated
by the vector field $\xi^\mu$, the metric changes as $\delta g_{\mu
  \nu} =
\nabla_{(\mu} \xi_{\nu)}$. So the integrated vertex operator that generates a gauge transformation for the gauge boson is
\be 
\int d^2z \left(
j_{L,z}^a j_{L,\bar z}^m + j_{L,\bar z}^a j_{L,z}^m
\right) \partial_m \xi_a \,.
\ee
The vector field $\xi_a$ has an $SU(2)$ index, and depends only on the
$AdS_3$ coordinates.  Such a vertex operator can be BRST
non-trivial if the vector field $\xi$ does not
vanish fast enough at infinity. This is related to the fact that a state of
the form $Q_{BRST} |\phi \rangle$ is not BRST exact if the state
$|\phi \rangle$ does not belong to the Hilbert space.  Working at
first order in the fermionic currents, we can rewrite the vertex
operator as:
\be  \int d^2z
(j_{L,z}^a \bar \partial \xi_a + j_{L,\bar z}^a \partial \xi_a).
\ee
We define the $n$-th mode of the boundary R-current $J^a_n$ as:
\be  
J^a_n= \int d^2z (j_{L,z}^a \bar \partial \xi^{(n)} + j_{L,\bar z}^a \partial \xi^{(n)}) \,,
\ee
where $\xi^{(n)} = \gamma^n+ \mathcal{O}(e^{-\phi})$ near the boundary
at $\phi \rightarrow \infty$.
We define the left-moving boundary R-current $J^a(x)$ as:
\be\label{defJR} J^a(x) = \sum_{n=-\infty}^{\infty} \frac{J^a_n}{x^{n+1}} \ee
where we have introduced the variable $x$ that parameterizes the eigenvalue of the parabolic generator of the symmetry group $SL(2,\mathbb{R})_L$. Asymptotically, the left-moving R-current therefore has the form
\be\label{defineJR}
J^a(x) = \int d^2 z \left[j^a_{L,z}\bar\p\left(\frac{1}{\gamma-x} + \mathcal{O}(e^{-\phi})\right) + j^a_{L,\zbar}\p\left(\frac{1}{\gamma-x} + \mathcal{O}(e^{-\phi})\right)\right]\,.
\ee
Similarly, we can also introduce the variable $\bar x$ related to
$SL(2,\mathbb{R})_R$. We can then define $\bar J^a(\bar x)$ using
diffeomorphisms with $\xi_{(n)} = \bar \gamma^n +
\mathcal{O}(e^{-\phi})$. The variables $(x,\bar x)$ can also be
interpreted to parameterize the manifold on which the spacetime
two-dimensional conformal field theory is defined.

Applying the diffeomorphism transformation to the action written out in detail in Appendix \ref{wsaction}, one can derive the explicit expression for the R-current in equation \eqref{defineJR}.

\subsubsection*{Non-trivial  Diffeomorphisms for the Interacting Theory}

In order to write down the exact expression for the R-current $J^a(x)$ in a convenient way we define the function $\Lambda(x,\bar x; \gamma, \bar \gamma,\phi)$, first introduced in \cite{Kutasov:1999xu}:
\be \Lambda(x,\bar x; \gamma, \bar \gamma,\phi) =
- \frac{1}{\gamma-x}\left[ \frac{(\gamma-x)(\bar \gamma - \bar x)e^{2 \phi}}{1+(\gamma-x)(\bar \gamma - \bar x)e^{2\phi}}\right]
\ee
We propose the following expression for the boundary R-current in the fully interacting theory: 
\be\label{JRdefn}
J^a(x) = \frac{1}{\pi} \int d^2 z 
( j_{L,z}^a \bar \partial \Lambda(x,\bar x;z,\bar z)  + j_{L,\bar z}^a \partial \Lambda(x,\bar x;z,\bar z)) \,.
\ee
Since $\Lambda$ behaves near the boundary as $\Lambda \approx
-\frac{1}{\gamma-x}$, it coincides with the expression derived in the
weak coupling, near-boundary region in \eqref{defineJR}. The
expression we put forward for the R-current is the natural
non-chiral generalization of the expressions of \cite{Kutasov:1999xu} for the vertex operators which generate the
asymptotic symmetry algebra.  Notice that a different choice for the
parameter $\xi^{(n)}$ would produce an operator related to $J^a(x)$ by a
trivial gauge transformation. The advantage of our choice of subleading behaviour is
that the function $\Lambda$ has a simple behaviour under the action of
the global symmetry group of the theory. Indeed, the function
$\Lambda$ satisfies the equation
\be \label{dxLambda}
\partial_{\bar x} \Lambda = \pi \Phi_1 \,, \ee
where the function $\Phi_1$ is defined as:
\be \Phi_1(x,\bar x;\gamma,\bar \gamma,\phi) = 
\frac{1}{\pi} \left( \frac{1}{(\gamma-x)(\bar \gamma - \bar x)e^\phi + e^{-\phi}} \right)^2 \,.\ee
This function is an eigenvector of the laplacian operator on euclidean
$AdS_3$, with zero eigenvalue. Near the boundary, this wave function
behaves like a delta-function identifying $\gamma$ and $x$. It is thus
the bulk-to-boundary propagator of a massless scalar from the boundary point $(x,\bar
x)$ to the bulk point $(\gamma, \bar \gamma,\phi)$.

In the worldsheet theory $\Phi_1$ is a primary field with respect to
the current algebra. It transforms in the discrete $\mathcal{D}_1^+
\times \mathcal{D}_1^+$ representation of the $SL(2,\mathbb{R})_L
\times SL(2,\mathbb{R})_R$ current algebra and with spin zero under
the action of the bosonic subgroup $SU(2)_L\times SU(2)_R$. We can
extend the representation to a representation of the full supergroup.

We wish to argue that the following equation holds true:
\be 
\label{delta}
\lim_{z\rightarrow w} \Phi_1(x,\bar x;z,\bar z)\Phi_1(y,\bar y;w,\bar w) 
=
 \delta^{(2)}(x-y)\Phi_1(y,\bar y;w,\bar w) + \mathcal{O}(z-w, \bar z - \bar w) 
\ee
The first justification for this is the matching of the worldsheet and
space-time conformal dimensions of the left and right hand
sides. Moreover, we recall that in the case of the Wess-Zumino-Witten
model, this equation was argued for on semi-classical grounds in
\cite{Kutasov:1999xu}, and was made precise in \cite{Teschner:1999ug}.
In \cite{Teschner:1999ug} the three-point function for fields in the
continuous representations in the Wess-Zumino-Witten model were
analytically continued to discrete values of the spin. In doing so,
one picks up residues of poles when shifting the contour of
integration over the spins that appear in the product of two operators
in the continuous representation.  These poles can arise from the
dynamical three-point function or from $SL(2,\mathbb{R})$
representation theory. What is important to us is that the
delta-function in equation (\ref{delta}) arose from $SL(2,\mathbb{R})$
group theoretic properties, namely from the analytic continuation in
the spin $j$ of the Clebsch-Gordan coefficient of the form
$|x|^{2j}$. Since this part of the three-point function is universal,
it is natural to assume that the delta-function appearing in the
product above is universal as well.

We collect some important formulae below that will prove to be very useful in our later calculations. We begin by noting that for $h=1$, equation \eqref{jLwithPhi} reads
\be
j_{L,z}(x;z) \Phi_1(y,\bar y; w,\bar w) = \frac{1}{z-w} \partial_y[(x-y)^2 \Phi_1(y,\bar y;w,\bar w)]
\ee
Moreover equation \eqref{dPrimary} reads, using the expression
\eqref{tDiffOps} for the $SL(2,\mathbb{R})$ generators in terms of
differential operators:
\be \partial_z \Phi_1(x,\bar x ;z ,\bar z) = -\frac{1}{c_+} \partial_x[:j_{L,z}\Phi_1:(x,\bar x;z,\bar z)]. \ee
If we assume $\Phi_1$ to be embedded into a primary of the current
algebra, 
this equation is true up to first order in the fermionic currents
as demonstrated in some detail
in appendix \ref{primaries}. Thus, from now on we work 
at leading order in the fermionic currents. 
Similarly we have the equation
\be\label{pbarPhiL} \partial_{\bar z} \Phi_1(x,\bar x ;z ,\bar z) = - \frac{1}{c_-} \partial_{\bar x}[:j_{L,\bar z}\Phi_1:(x,\bar x;z,\bar z)] \ee
Using the expression of the stress-tensor in terms of the
right-current, we obtain similar relations:
\be 
\partial_z \Phi_1(x,\bar x ;z ,\bar z) = -\frac{1}{c_-} \partial_x[:j_{R,z}\Phi_1:(x,\bar x;z,\bar z)] \ee
\be \partial_{\bar z} \Phi_1(x,\bar x ;z ,\bar z) = - \frac{1}{c_+} \partial_{\bar x}[:j_{R,\bar z}\Phi_1:(x,\bar x;z,\bar z)]. 
\ee
{From} these relations and by integrating over $\bar x$, we deduce the following equations satisfied by the operator $\Lambda$:
\begin{align} \label{dLambda}
\partial_z \Lambda(x,\bar x;z,\bar z) &= -\frac{\pi}{c_-} :j_{R,z} \Phi_1:(x,\bar x;z,\bar z) \cr
\partial_{\bar z} \Lambda(x,\bar x;z,\bar z) &= -\frac{\pi}{c_+} :j_{R,\bar z} \Phi_1:(x,\bar x;z,\bar z) \,.
\end{align}
All these relations will be repeatedly used in the derivation of the algebra of currents in the boundary theory.

\subsection{Computation of the Spacetime R-Current Algebra}

In order to show that the operator $J^a(x)$ defined in equation
\eqref{JRdefn} is the R-current of the spacetime CFT, we want to
compute the spacetime operator product singularity inside correlation
functions in the following limit:
\be 
\lim_{x \to y} J^a(x) \cdot J^b(y)\,. 
\ee
However it turns out to be more convenient to compute an OPE involving an anti-holomorphic derivative of one of the currents, namely  
\be 
\lim_{x \to y} \partial_{\bar x} J^a(x) \cdot J^b(y) \,,
\ee
and to integrate the result with respect to $\bar x$. We will therefore compute the OPE between the current 
\be 
J^b(y) =  -\frac{1}{\pi}\int d^2 w\left[ j_{L,\bar z}^b\partial_w\Lambda(y,\bar y;w,\bar w)+j_{L,z}^b\partial_{\bar w}\Lambda(y,\bar y; w,\bar w)\right]\,,
\ee
and its derivative with respect to $\bar x$, which, using equation \eqref{dxLambda}, can be written as
\begin{align} 
\partial_{\bar x} J^a(x) =  - \int d^2 z \left[ j_{L,\bar z}^a \partial_{z}\Phi_1(x,\bar x;z,\bar z) + j_{L, z}^a\partial_{\zbar}\Phi_1(x,\bar x;z,\bar z) \right]
\end{align}
where $a$ and $b$ are $SU(2)$ indices.  Following
\cite{Kutasov:1999xu}, we regularize by cutting small holes in the
worldsheet, at the points where operators are inserted. This implies
that we can freely use the equation of motion (that have
contact terms singularities with the other operators on the worldsheet),
 but the integration by parts gives rise to boundary
terms\footnote{We can also work with a worldsheet without holes. In
  that case we can integrate by parts freely, but the contact terms between 
the equations of motion and the other operators will contribute. }.  We can think of the operators $J^b(y)$
and $\partial_{\bar x} J^a(x)$ as being inserted within a worldsheet
correlation function. Since we are interested in the (spacetime) OPE
between these two operators, we will only keep track of the terms
arising when these operators are close one to another on the
worldsheet, and discard the possible contribution due to the presence
of other operators in the correlation function.
A general justification for this procedure can be found in
\cite{Aharony:2007rq}. So we will write the $\bar x-$derivative of
the R-current as 
\be
\partial_{\bar x} J^a(x) =   \frac{1}{i}\left[\oint_w d\bar z j^a_{L,\zbar}\Phi_1(x,\bar x;z,\bar z) + \oint_w dz j^a_{L,z}\Phi_1(x,\bar x;z,\bar z) \right] \,,
\ee
where the contour integral runs over the boundary of the small disc cut out around the position of the integrated operator in $J^b(y)$. The OPE we wish to compute is
\begin{align}\label{pJJ1}
\partial_{\bar x} J^a(x) \cdot J^b(y)
= - \frac{1}{ \pi i} \int d^2 w &\left[ \left(
j_{L,\bar z}^b\partial_w\Lambda(y,\bar y;w,\bar w)+j_{L,z}^b\partial_{\bar w}\Lambda(y,\bar y;w,\bar w)
\right) \right. \cr
& \hspace{1in} \left.
\cdot \left(
\oint_w d\bar z j^a_{L,\zbar}\Phi_1 + \oint_w dz j^a_{L,z}\Phi_1
\right) \right]
\end{align}
The contour integrals around the point $w$ will pick up the singular
terms in the OPE between the integrated composite
operators. First notice that there are no singular terms arising between the
$SU(2)$ currents and the operators $\Phi_1$, $\partial_w\Lambda$ and
$\partial_{\bar w}\Lambda$ since the later transform in the trivial
representation under the action of $SU(2)$. Moreover there is no
singularity either in the OPE between $\Phi_1$ and
$\partial_w\Lambda$, $\partial_{\bar w}\Lambda$. To prove this we use
the formula \eqref{dLambda}:
\begin{align}\label{PhipLambda}
\lim_{z \to w} \Phi_1(x,\bar x ; z, \bar z) \partial_w\Lambda(y,\bar y;w,\bar w)
&= -\frac{\pi}{c_-} \lim_{z \to w} \Phi_1(x,\bar x ; z, \bar z) :j_{R,z} \Phi_1:(y,\bar y;w,\bar w) \cr
&= - \frac{\pi}{c_-} \frac{c_-}{c_++c_-}\frac{\partial_{\bar x}[(\bar x-\bar y)^2 \delta^{(2)}(x-y)\Phi_1(y,\bar y;w,\bar w)] }{z-w} \cr
 &\hspace{1in} - \frac{\pi}{c_-} \delta^{(2)}(x-y):j_{R,z}\Phi_1:(y,\bar y;w,\bar w)\cr
&\hspace{1.5in}+ \mathcal{O}(z-w, \bar z - \bar w) \cr
&=  \delta^{(2)}(x-y)  \partial_w\Lambda(y,\bar y;w,\bar w) + \mathcal{O}(z-w, \bar z - \bar w)\,.
\end{align}
The previous computation is straightforwardly generalized to the OPE between 
the operators $\Phi_1$ and $\partial_{\bar w}\Lambda$. We conclude that the only singular terms picked up by the contour integral in 
equation \eqref{pJJ1} come from the OPE between two $SU(2)$ currents.
We obtain:
\begin{align}
 \partial_{\bar x} J_R^a(x) &J_R^b(y) =  -\frac{1}{2 \pi i} \int d^2 w  \cr
&\times \left\{\oint_w d \bar z \left(
\frac{\kappa^{ab}c_3}{(\bar z - \bar w)^2} + f^{ab}{}_c c_4 \frac{j^c_{\bar z}(w)}{\bar z - \bar w}
+ f^{ab}{}_c (c_4-g) \frac{j^c_{z}(w)(z-w)}{(\bar z - \bar w)^2}
\right) \right. \cr
& \hspace{2in}\times 
\left(\lim_{:z \to w:} \Phi_1(z,\bar z; x,\bar x)  \partial_w\Lambda(w,\bar w;y,\bar y) \right) \cr
& +\oint_w d \bar z \left(
\tilde{c} \kappa^{ab} 2\pi \delta^{(2)}(z-w) + f^{ab}{}_c \frac{(c_4-g)j^c_z(w)}{\bar z - \bar w}
+  f^{ab}{}_c \frac{(c_2-g)j^c_{\bar z}(w)}{ z - w}\right) \cr 
&  \hspace{2in}\times \left(\lim_{:z \to w:} \Phi_1(z,\bar z; x,\bar x)  \partial_w\Lambda(w,\bar w;y,\bar y) \right)  \cr
& +\oint_w d  z \left(
\tilde{c} \kappa^{ab} 2\pi \delta^{(2)}(z-w) + f^{ab}{}_c \frac{(c_4-g)j^c_z(w)}{\bar z - \bar w}
+  f^{ab}{}_c \frac{(c_2-g)j^c_{\bar z}(w)}{ z - w}\right) \cr
&  \hspace{2in}\times \left(\lim_{:z \to w:} \Phi_1(z,\bar z; x,\bar x)  \partial_{\bar w} \Lambda(w,\bar w;y,\bar y) \right)  \cr
& + \oint_w d  z \left(
\frac{\kappa^{ab}c_1}{( z - w)^2} + f^{ab}{}_c c_2 \frac{j^c_{ z}(w)}{ z -  w}
+ f^{ab}{}_c (c_2-g) \frac{j^c_{\bar z}(w)(\bar z-\bar w)}{( z - w)^2}\right) \cr
&  \hspace{2in}\times \left.  \left(\lim_{:z \to w:} \Phi_1(z,\bar z; x,\bar x)  \partial_{\bar w} \Lambda(w,\bar w;y,\bar y) \right)   \right\}\,.
\end{align}
We get twelve terms (three on each double-line) that we denote $A_1\,,\ldots, A_{12}$. 
We can now explicitly perform the contour integrals. 
One observes that the terms $A_3$, $A_6$, $A_8$ and $A_{12}$ vanish. The contour integrals in $A_2$, $A_9$, $A_5$ and $A_{11}$ can be simply performed. The regular limit $\lim_{:z \to w:} \Phi_1(z,\bar z; x,\bar x)  \partial_w\Lambda(w,\bar w;y,\bar y)$ can then be read from equation \eqref{PhipLambda} (and similarly for the limit involving the anti-holomorphic derivative of $\Lambda$). 
Using the relation $c_2+c_4-g=i$ from 
equations
\eqref{relationsCs}, the terms  $(A_2+ A_9)$ and $(A_5+ A_{11})$ can be simplified separately and these four terms combine to give 
\be
A_2+A_9 + A_5 + A_{11} = 2\pi \delta^{(2)}(x-y)\, i{f^{ab}}_c J^c(y) \,.
\ee
This leaves us with computing the terms $A_1, A_4, A_7, A_{10}$, which involve double poles. We deal with $A_1$ explicitly first:
\begin{align}
A_1 &=  -\frac{1}{\pi i} \int d^2 w \oint_w d \bar z \frac{\kappa^{ab}c_3}{(\bar z - \bar w)^2} 
\left[\lim_{:z \to w:} \Phi_1(x,\bar x; z,\bar z)  \partial_w\Lambda(y,\bar y;w,\bar w) \right] \cr
&= -2\kappa^{ab}c_3 \int d^2 w \left[
\lim_{:z \to w:} \partial_{\bar z} \Phi_1(x,\bar x; z,\bar z)  \partial_w\Lambda(y,\bar y;w,\bar w)
\right] \cr
&=  \frac{2 \kappa^{ab}c_3}{c_-} \int d^2 w \p_x\left[ 
\lim_{:z \to w:} :j_{L,\bar z} \Phi_1:(x,\bar x; z,\bar z)  \partial_w\Lambda(y,\bar y;w,\bar w)
\right] \,.
\end{align}
where we performed the contour integral and replaced the derivative of $\Phi_1$ using the formula \eqref{pbarPhiL}
We have to compute the regular term in the OPE between $j_{L,\bar z} \Phi_1:(z,\bar z; x,\bar x)$ and $\partial_w\Lambda(w,\bar w;y,\bar y)$. Using formula \eqref{dLambda}, we have:
\begin{align}
\lim_{:z \to w:}& :j_{L,\bar z} \Phi_1:(x,\bar x; z,\bar z)  \partial_w\Lambda(y,\bar y;w,\bar w) \cr
&= -\frac{\pi}{c_-} \lim_{:z \to w:} :j_{L,\bar z} \Phi_1:(x,\bar x; z,\bar z)  :j_{R,z}\Phi_1:(y,\bar y;w,\bar w) \cr
&= -\frac{\pi}{c_-} \delta^{(2)}(x-y)  :j_{L,\bar z}:j_{R,z}\Phi_1::(y,\bar y;w,\bar w) \cr
& = \delta^{(2)}(x-y)  :j_{L,\bar z} \partial_w\Lambda:(y,\bar y;w,\bar w) 
\end{align}
We deduce that
\be 
A_1 =2 \kappa^{ab}\frac{c_3}{c_-} \partial_x \delta^{(2)}(x-y)\int d^2 w :j_{L,\bar z} \p_w\Lambda:(y,\bar y; w,\bar w)\,. 
\ee
A similar analysis for the double pole term in $A_{10}$ leads to 
\be
A_{10} = 2\kappa^{ab}\frac{c_1}{c_+} \pi \partial_x \delta^{(2)}(x-y)\int d^2 w :j_{L,z}\p_{\wbar}\Lambda:(y,\bar y; w,\bar w).
\ee
The contact terms in $A_4$ and $A_7$ contribute due to the following integrals:
\begin{align} 
\oint_w dz \delta^{(2)}(z-w) f(z,\zbar) = \frac{-1}{2\pi} \oint_w dz \partial_{\bar w} \frac{1}{z-w} f(z,\zbar)
&= -i \partial_{\bar w} f(w,\wbar)\cr
\oint_w d\zbar \delta^{(2)}(z-w) f(z,\zbar) = \frac{-1}{2\pi} \oint_w dz \partial_{w} \frac{1}{\zbar-\wbar} f(z,\zbar)
&= -i \partial_{w} f(w,\wbar)\,. 
\end{align}
Using these results and following the steps we did for the evaluation
of $A_1$, we find that
\begin{align}
A_4 &= 2\kappa^{ab}\tilde c\int d^2 w \left[ \lim_{:z\rightarrow w:}
\p_{\zbar}\Phi_1(x,\bar x; z,\bar z)\p_w\Lambda(y,\bar y;w,\bar w)\right] \cr
&= -2\kappa^{ab}\frac{\tilde{c}}{c_-}\p_{x}\delta^{(2)}(x-y) \int d^2 w :j_{L,\zbar}\p_w\Lambda:(y,\bar y;w,\bar w)\cr
\text{and}\quad A_7 &=-2\kappa^{ab}\frac{\tilde{c}}{c_+}\p_{x}\delta^{(2)}(x-y) \int d^2 w :j_{L,z}\p_{\wbar}\Lambda:(y,\bar y;w,\bar w) \,.
\end{align}
Combining these four terms and using the relations between coefficients \eqref{relationsCs}, we get 
\be
A_1+A_4+A_7+A_{10} = -2\pi\kappa^{ab}\p_{x}\delta^{(2)}(x-y)\, I(y,\ybar) \,,
\ee
where we have introduced the central extension operator $I$ of the R-current algebra:
\be\label{1stDefI}
I(y,\bar y) = \frac{1}{\pi}\int d^2w \left[j_{L,w}\p_{\wbar}\Lambda(y,\bar y;w,\bar w) + j_{L,\wbar}\p_w\Lambda(y,\bar y;w,\bar w)\right]\,.
\ee
We shall turn to the study of this operator after deriving the algebra
involving the other bosonic currents on the boundary. For now, we
observe that when $f^2=1/k$, one can check that this reduces to the
operator $I$ defined in \cite{Kutasov:1999xu}\footnote{ Our definition
  of $I$ differs from the one of \cite{Kutasov:1999xu} by an overall
  normalization factor of
  $k$}. 
We can also check that the operator $I(y,\bar y)$ does not depend on
the spacetime coordinates $y,\bar y$:
\begin{align}\label{ddbarI}
\partial_{\bar y} I(y,\bar y) &= 
 \int d^2w \left[j_{L,w}\p_{\wbar}\Phi_1(y,\bar y;w,\bar w) + j_{L,\wbar}\p_w\Phi_1(y,\bar y;w,\bar w)\right] \cr
& =  \frac{1}{i} \oint dw \left[j_{L,w}\Phi_1(y,\bar y;w,\bar w)\right] - \frac{1}{i} \oint d\bar w \left[ j_{L,\wbar}\Phi_1(y,\bar y;w,\bar w)\right] \cr
& = - \frac{c_+}{\pi i} \oint dw \left[\p_w \bar \Lambda(y,\bar y;w,\bar w)\right] - \frac{c_-}{\pi i} \oint d\bar w \left[\p_{\bar w} \bar \Lambda(y,\bar y;w,\bar w)\right]\cr
&=0\,.
\end{align}
From the first line to the second, we integrated by parts and used the current conservation. From the second to the third, we used the (spacetime) complex conjugate of equation \eqref{dLambda} that involves the operator $\bar \Lambda$ defined as: $\p_x \bar \Lambda = \Phi_1$. Similarly we can prove that $\partial_{ y} I(y,\bar y)=0$.

Putting together what we have so far, we find the OPE
\be
\p_{\xbar}J^a(x)\cdot J^b(y) \sim -2\pi\kappa^{ab} \p_{x}\delta^{(2)}(x-y)\, I + 2\pi \delta^{(2)}(x-y)\, i {f^{ab}}_c J^a(y) \,,
\ee
We can integrate with respect to $\bar x$ and find:
\be 
J^a(x)\cdot J^b(y) \sim \kappa^{ab} \frac{1}{(x-y)^2}\, I + \frac{1}{x-y}\, i {f^{ab}}_c J^c(y) \,.
\ee
We observe that the $SU(2)_R$ symmetry of the $N=4$ superconformal
algebra is at level $I$.  By the structure of the $N=4$
superconformal algebra, this implies a central charge equal to
$c=6 \, I$. We will confirm the value of the central charge by
evaluating the operator product expansion of the stress-tensor with itself.

\section{The Spacetime Virasoro Algebra}\label{Virasoro}

In the previous section we have identified the vertex operators that correspond, in spacetime, to the 
(left) R-currents $J^a(x)$. They are part of the generators of the small $N=(4,4)$
superconformal symmetry of the holographically (boundary) dual conformal field theory. In this
section we will address the construction of the full set of
left generators. The generators of the space-time right-moving superconformal algebra can be 
constructed similarly.
We put forward the following expression for the vertex operator of the boundary stress tensor:
\begin{align}\label{defT}
T(x) = -\frac{1}{2\pi}\int d^2z&\left[ (\p_x j_{L,z} \p_x \p_{\bar z} \Lambda(x,\bar x;z,\bar z)+2\p_x^2 j_{L,z}\p_{\bar z} \Lambda(x,\bar x;z,\bar z) )
\right. \cr
& \left. + (\p_x j_{L,\zbar} \p_x \p_z \Lambda(x,\bar x;z,\bar z) + 2 \p_x^2 j_{L,\zbar}\p_z \Lambda(x,\bar x;z,\bar z)) \right]  \,.
\end{align}
Once again, this is a non-chiral generalization of the expression for
the stress tensor for the case with pure NS flux \cite{Kutasov:1999xu}.  First we will
compute the OPE between this operator and the R-current
\eqref{JRdefn}.  We shall show that, given our basic OPEs of the
worldsheet currents and primary operators, this reproduces the
superconformal OPEs of the boundary operators. 

Before we delve into this calculation, it is important to note that
the operator is closed in the hybrid cohomology. The operator is
independent of the ghosts and therefore it is BRST closed when it is a
worldsheet conformal primary of conformal dimension one -- this
follows from formula (\ref{BRSTcurrent}). Using the operator product
expansion between the worldsheet energy-momentum tensor, the currents
and the field $\Lambda$, it is possible to show that this is true for
the particular coefficients chosen in equation (\ref{defT}). 
One can alternatively fix the coefficients in
formula (\ref{defT}) by demanding that the energy-momentum tensor
transform as a weight $2$ tensor under the global $SL(2,\mathbb{R})_L$
symmetry group (up to a
BRST exact operator). As a BRST-closed operator, we can insert it in hybrid
superstring amplitudes, and compute Ward identities for the resulting
correlator.

As for the previous computation, it is simpler to consider the OPE between the $\bar x$-derivative of the current and the stress-energy tensor. We wish to evaluate the OPE:
\begin{align}\label{JTstep1}
\p_{\bar x} J^a(x)\cdot T(y) = & - \frac{1}{2\pi i}\left[\oint_w dz j^a_{L,z}\Phi_1(x,\bar x;z,\bar z) + \oint_w d\zbar j^a_{L,\bar z}\Phi_1(x,\bar x;z,\bar z)\right] \cr
&\hspace{.4in} \times \left[ \int d^2w\left[ (\p_y j_{L,z} \p_y \p_{\bar w} \Lambda(y,\bar y;w,\bar w)+2\p_y^2 j_{L,z}\p_{\bar w} \Lambda(y,\bar y;w,\bar w) ) \right.\right. \cr
&\hspace{.8in}\left.\left.  + (\p_y j_{L,\zbar} \p_y \p_w \Lambda(y,\bar y;w,\bar w) + 2 \p_y^2 j_{L,\zbar}\p_w \Lambda(y,\bar y;w,\bar w)) \right]\right].
\end{align}
The contour integral will pick up the poles in the OPEs between the
integrated operators. The $SU(2)$ currents $j^a_{L,\zbar}$ and
$j^a_{L,z}$ will not contribute to these poles since all of the
integrated operators in the stress-tensor transform trivially under
the $SU(2)_L$ symmetry. So the singular terms only come from the OPE
between the operator $\Phi_1$ and the integrated operators in the
stress-tensor. Notice that there is no short-distance singularity
between $\Phi_1$ and $\p_z \Lambda$. Indeed using equation
\eqref{dLambda} we can rewrite the later as $:j_{R,z}\Phi_1:$. The
right current has a pole in its OPE with $\Phi_1$, but the coefficient
cancels against the delta-function appearing when the two $\Phi_1$
operators come close to each other. The same is true for $\p_{\bar z}
\Lambda$. We conclude that the only singular terms that will
contribute to the OPE \eqref{JTstep1} come from the short-distance
singularity between the $\Phi_1$ operators integrated in the derived
R-current, and the $SL(2,\mathbb{R})$-left currents integrated in the
stress-tensor.

We can expand the OPE \eqref{JTstep1} as a sum of eight
terms. Consider, for instance, the term obtained by taking the OPE
between the first terms in each of the bracketed terms in
\eqref{JTstep1}.
As argued before, the contour integral picks up the pole coming from the short distance
singularity between the operator $\Phi_1$ and the $SL(2,\mathbb{R})$
current:
\begin{align}
B_1 &= -\frac{1}{2\pi i}\int d^2w \left[
\oint_w dz j^a_{L,\zbar}\Phi_1(x,\bar x;z,\bar z) \p_y j_{L,z} \p_y \p_w \Lambda(y,\bar y;w,\bar w) \right] \cr
& = -\frac{1}{2\pi i}\int d^2w \left[ \oint_w dz j^a_{L,\zbar} \lim_{:z\to w:} 
\p_y \left( \frac{\p_x[(x-y)^2 \Phi_1(x,\bar x;z,\bar z)]}{w-z}\right)\p_y \p_w \Lambda(y,\bar y;w,\bar w) \right] \cr
& =  \int d^2w  \left[ j^a_{L,\zbar} \lim_{:z\to w:} \p_x [2(y-x)\Phi_1(x,\bar x;z,\bar z) \p_y \p_w \Lambda(y,\bar y;w,\bar w) ]   \right] \cr
& = - 2 \p_x \delta^{(2)}(x-y) \int d^2w  \left[ j^a_{L,\zbar} \p_w \Lambda(y,\bar y;w,\bar w)\right] \,.
\end{align}
Using the same techniques, we can compute all of the remaining terms.
We suppress the details since the calculations are very similar to the
ones we have already performed. Combining these terms, we eventually
obtain the spacetime OPE
\be
\p_{\bar x} J^a(x) T(y) \sim -2\pi \p_x \delta^{(2)}(x-y) J^a(y) \,,
\ee
which can be integrated with respect to $\bar x$ to give:
\be
J^a(x) T(y)\sim  \frac{J^a(y)}{(x-y)^2}.
\ee
By evaluating the right-hand side at the point $x$, we get that the current
is a conformal primary of dimension one in space-time:
\be
 T(y) J^a(x) \sim \frac{J^a(x)}{(x-y)^2} +  \frac{\partial_x J^a(x)}{y-x}.
\ee
One can also check that with our definition, the stress-tensor satisfies the OPE
that codes the Virasoro algebra:
\be\label{TTOPEeqn}
T(x)\cdot T(y) \sim \frac{3I}{(x-y)^4} + \frac{2T(y)}{(x-y)^2}+ \frac{\p T(y)}{(x-y)} \,,
\ee
which gives for the central charge $c=6 \, I$, where $I$ is the
central extension. The details are given in the Appendix \ref{TTOPE}
and confirms the value of the central charge obtained using the
$R$-current algebra. This is a consistency check on our
proposal for the vertex operators of the boundary currents. In the
following, we will discuss some of the properties of the central
extension in more detail. But first, we briefly outline how our
discussion may be extended to also describe the superconformal
generators.

\subsection{Superconformal Generators}

With respect to the scaling operator, we have the $sl(2,r)$ generators 
of charges $\pm 1, 0$, and the $su(2)$ charges which are inert. The eight
fermionic generators have charges $\pm \frac{1}{2}$. It is natural then
to gather the eight fermionic currents into four $x$-dependent currents:
\begin{eqnarray}
j_F^{\tilde{a}\alpha}(x) &=& e^{-x {\cal J}_0^-} j_F^{+\frac{1}{2},\tilde{a}\alpha}
 e^{+x {\cal J}_0^-}
\nonumber \\
&=&  j_F^{+\frac{1}{2},\tilde{a}\alpha} - x  j_F^{-\frac{1}{2},\tilde{a}\alpha},
\end{eqnarray}
where $\tilde{a}$ indicates an $su(2)_R$ doublet.  The current must have weight $(-\frac{1}{2},0)$ in space-time, therefore, the superconformal generators should be of the general form
\begin{eqnarray}
G^{\tilde{a}\alpha} (x)
&=& \int_{\Sigma} b_1 j_F^{\tilde{a}\alpha} (x) \otimes j_R \partial_x \Phi_1 
+ b_2 \partial_x  j_F^{\tilde{a}\alpha} (x) \otimes j_R \Phi_1.
\end{eqnarray}
These operators have weight $(\frac{3}{2},0)$ in space-time. Imposing
that the operator transforms appropriately with respect to the raising operator
in the $sl(2, \mathbb{R})$ algebra gives rise to the
relation
\begin{eqnarray}
b_2 = 2 b_1.
\end{eqnarray}
We thus propose the following operator to generate the supersymmetry transformations of the boundary theory:
\begin{eqnarray}
G^{\tilde{a}\alpha} (x)
&=& b \int_{\Sigma} j_F^{\tilde{a}\alpha} (x) \otimes j_R \partial_x \Phi_1 
+ 2 \partial_x  j_F^{\tilde{a}\alpha} (x) \otimes j_R \Phi_1.
\end{eqnarray}
This is the analogue of the expressions in \cite{Kutasov:1999xu} in the zero ghost picture.

\subsection*{Higher superfield components}

In order to prove that we have an $N=4$ superconformal algebra in
space-time, we should proceed along the lines of the calculation of
the R-current algebra and compute the OPE of
$\partial_{\bar{x}}G^{\tilde{a}\alpha} (x)\cdot G^{\tilde{b}\beta}
(y)$ within a correlation function. There are some qualitatively new
features that arise when we attempt to compute it, such as the OPE
between the fermionic currents and the operator $\Phi_1$.

This calculation can guide us in lifting the limitations on our
derivation of the R-current and Virasoro algebras. The vertex
operators for the physical fields described in \cite{BVW} have a full
supermultiplet worth of fields and should be thought of as superfields
in spacetime (see e.g. \cite{Gotz:2005ka, Gotz:2006qp}). However,
as is clear from our expressions of the vertex operators in equations
\eqref{JRdefn} and \eqref{defT}, we have chosen to truncate to a
bosonic component of the superfield in spacetime. To compute the super
conformal algebra in space-time we need to take into account
the first fermionic correction to these superfields, and we need to
work at least to first order in the fermionic currents.

We leave this for future work and proceed to discuss
properties of the central extension operator.

\subsection{Further remarks}\label{centralextension}

{From} the operator product expansion involving the boundary stress
tensor and the R-currents, we found that the central charge of the
boundary theory is given by the operator $6 \, I$, where $I$ is the
operator
\be\label{defI}
I = \frac{1}{\pi}\int d^2w \left[j_{L,w}(w;x)\p_{\wbar}\Lambda(w;x) + j_{L,\wbar}(w;x)\p_{w}\Lambda(w;x)\right] \,.
\ee
We would like to make a few observations regarding the central
extension operator.

\begin{enumerate}

\item The operator $I(x,\xbar)$ is independent of the spacetime
  coordinates $(x,\xbar)$. The proof for this has been given in
  equation \eqref{ddbarI}. Thus it behaves like a constant in 
correlation functions.

\item Near the boundary of $AdS$ space, in the $\phi\rightarrow
  \infty$ limit, the operator $I$ takes the form
\be
I =  -\int d^2 z ( c_+\pbar\gamma \p\bar\gamma +c_-\p\gamma\pbar\bar\gamma)\delta^{(2)}(x-\gamma) - \frac{1}{\pi} \int d^2z e^{2\phi} (c_+\p\bar\gamma\pbar\gamma + c_-\p\gamma \pbar\bar\gamma) + \ldots 
\ee
{From} the form of the vertex operator we can see that it is an
off-diagonal mode of the metric (mixed with an anti-symmetric tensor).

\item For the case of pure NSNS flux, the operator $I$ that we found
  agrees with the central extension operator found in
  \cite{Kutasov:1999xu}\footnote{Our definition differs from the one
    in \cite{Kutasov:1999xu} by an overall normalization factor of
    $1/k$.}. We refer the reader to \cite{Giveon:2001up} for a
  detailed analysis regarding the behaviour and the interpretation of
  the operator $I$ in correlation functions.
\end{enumerate}
Note in particular that by the fact that the operator $6 \, I$ appears
as the coefficient of the boundary two-point function for the boundary
energy-momentum tensor, we know that it governs the value of the
conformal anomaly. In a background $AdS_3$ space-time, there are
various other ways to compute the value of the conformal anomaly
coefficient in the supergravity approximation, namely from the
Virasoro algebra directly \cite{Brown:1986nw} or from the holographic
Weyl anomaly  \cite{Henningson:1998gx}. These imply that in
an $AdS_3$ background the vacuum expectation value of the central
extension operator is equal to the Brown-Henneaux central charge as
determined in supergravity. This has been argued in more detail in
\cite{de Boer:1998pp, Kutasov:1999xu, Giveon:2001up} for the NS-NS case.
Our calculation
is also valid in the strongly curved regime, and codes corrections to
the central extension operator that can occur in excited states 
(as in \cite{Kutasov:1999xu, Giveon:2001up}). 

\section{Summary and Conclusions}\label{conclusions}

Our main result is the construction of the vertex operators for the
boundary superconformal algebra, in the worldsheet theory that
describes string theory on $AdS_3\times S^3$ with both Ramond-Ramond
and Neveu-Schwarz-Neveu-Schwarz fluxes. We have verified that they
satisfy the conformal operator algebra using stringy worldsheet
techniques. A crucial technical ingredient was the
conformal current algebra \cite{ABT} satisfied by the
currents in the supergroup model $PSU(1,1|2)$. It is important to note
that most of the simplifications that arise in the calculations follow
from constraints on  the coefficients appearing in the
current algebra. These constraints, in turn, follow from the
existence of a supergroup valued field throughout the
two-dimensional moduli space
parameterized by the fluxes.

It is a good check on our construction that at Wess-Zumino-Witten
points (where the coefficient of the kinetic term is equal in absolute
value to the coefficient of the Wess-Zumino term), which is the case
with only NS-NS flux, our results coincide with those obtained in the
NSR formalism in \cite{de Boer:1998pp, Kutasov:1999xu,
  Giveon:2001up}. Technically, our results are a non-chiral
generalization of those references. Our generalization includes
backgrounds which are near-brane limits of D1-D5 D-brane systems for
which the infrared limit of the dual gauge theory in the Higgs phase
can be studied directly.  An interesting result is the explicit
expression for the central extension operator.

There are many directions for future work.
A clear challenge is to extend our analysis to the full $N=4$ superconformal algebra.
A related problem is to further exploit the $PSU(1,1|2)$ superisometries
of the hybrid formalism to covariantize further our analysis. One can also 
attempt to generalize our arguments to other $AdS_3$ backgrounds that
can be described in a Berkovits type formalism.

We would like to argue that our construction is a significant
first step in the construction of the full spectrum of string theory
(including black hole excitations) in the $AdS_3$ background with
Ramond-Ramond fluxes, with appropriate boundary conditions. Indeed,
the full local asymptotic symmetry algebra in space-time will be the
natural tool to classify the spectrum, thus reducing the spectral
problem to listing primary states. The latter often reduces to a
mini-superspace problem.

Our construction of the spacetime symmetry generators may be extended
to $AdS_2 \times S^2$ backgrounds. String theory in $AdS_2 \times S^2$
with Ramond-Ramond fluxes can be described using the four-dimensional
hybrid formalism \cite{Berkovits:1994wr}. The target space in that case
is the geometric coset $PSU(1,1|2)/(U(1) \times U(1))$
\cite{Berkovits:1999zq}. Since the gauging involves only the right
symmetry, a reasonable hypothesis is that the left-current algebra
will be left intact. Then our previous construction will extend
straightforwardly. Notice that we obtain a set of holomorphic
generators in spacetime, as suited for a chiral two-dimensional CFT
dual to quantum gravity on an $AdS_2$ spacetime.

Finally, the conformal current algebra of the supergroup sigma-model
is tied in with both the worldsheet integrable structure of the
sigma-model, and the infinite set of conserved charges in
space-time. It would be useful to understand better the relations
between these two important features of the theory.

\section*{Acknowledgements}

We would like to thank Costas Bachas, Denis Bernard, Jaume Gomis,
Robert Myers and Amit Sever for helpful discussions.

\begin{appendix}

\section{Worldsheet Action for Strings on $AdS_3$ with RR Flux}\label{wsaction}

\subsection{Matrix Generators for $PSU(1,1|2)$}

We will now give an explicit matrix representation of the superalgebra
which we will use to write down the worldsheet action. Recall that the
Pauli sigma-matrices are given by
\begin{align} 
\sigma_1 = \left( \begin{array}{cc} 0&1 \\ 1&0 \end{array} \right) \qquad
\sigma_2 = \left( \begin{array}{cc} 0&-i \\ i&0 \end{array} \right) \qquad
\sigma_3 = \left( \begin{array}{cc} 1&0 \\ 0&-1 \end{array} \right) \,.
\end{align}
They satisfy the algebra
\be \sigma_a \sigma_b = \delta_{ab}I + i\epsilon_{abc}\sigma_c \,. \ee
Then we can write the bosonic generators $K_{ab}$ in terms of the Pauli matrices:
\beq K_{12} = \left( \begin{array}{cc} -\frac{i}{2}\sigma_3 &0 \\ 0&-\frac{i}{2}\sigma_3 \end{array} \right) & 
K_{13} = \left( \begin{array}{cc} \frac{i}{2}\sigma_2 &0 \\ 0& \frac{i}{2}\sigma_2 \end{array} \right) &
K_{23} = \left( \begin{array}{cc} -\frac{i}{2}\sigma_1 &0 \\ 0&-\frac{i}{2}\sigma_1 \end{array} \right) \\
K_{14} = \left( \begin{array}{cc} -\frac{i}{2}\sigma_1 &0 \\ 0&\frac{i}{2}\sigma_1 \end{array} \right) & 
K_{24} = \left( \begin{array}{cc} -\frac{i}{2}\sigma_2 &0 \\ 0&\frac{i}{2}\sigma_2 \end{array} \right) &
K_{34} = \left( \begin{array}{cc} -\frac{i}{2}\sigma_3 &0 \\ 0&\frac{i}{2}\sigma_3 \end{array} \right) \,.
\eeq
Similarly, we represent the fermionic generators as  
\be \begin{array}{ll} S_{11} = \left( \begin{array}{cc} 0 & \frac{1}{2}\sigma_1  \\ -\frac{1}{2}\sigma_1 & 0 \end{array} \right) & 
S_{21} = \left( \begin{array}{cc} 0 & \frac{1}{2}\sigma_2  \\ -\frac{1}{2}\sigma_2 & 0 \end{array} \right) \nonumber \\
S_{31} = \left( \begin{array}{cc} 0 & \frac{1}{2}\sigma_3  \\ -\frac{1}{2}\sigma_3 & 0 \end{array} \right) & 
S_{41} = \left( \begin{array}{cc} 0 & -\frac{i}{2} \mathbb{I}  \\ -\frac{i}{2}\mathbb{I} & 0 \end{array} \right) \nonumber \\
S_{12} = \left( \begin{array}{cc} 0 & -\frac{1}{2}\sigma_1  \\ -\frac{1}{2}\sigma_1 & 0 \end{array} \right) & 
S_{22} = \left( \begin{array}{cc} 0 & -\frac{1}{2}\sigma_2  \\ -\frac{1}{2}\sigma_2 & 0 \end{array} \right) \nonumber \\
S_{32} = \left( \begin{array}{cc} 0 & -\frac{1}{2}\sigma_3  \\ -\frac{1}{2}\sigma_3 & 0 \end{array} \right) & 
S_{42} = \left( \begin{array}{cc} 0 & \frac{i}{2} \mathbb{I}  \\ -\frac{i}{2}\mathbb{I} & 0 \end{array} \right) \,.
\end{array}
\ee 
They give the generators in the fundamental representation of
$sl(2|2)$. The generators $S$ square to a multiple of
the identity. We choose the invariant metric: 
\be \langle K_{ab},
K_{cd} \rangle = -\epsilon_{abcd} = Str(K_{ab} K_{cd}) \qquad \langle
S_{a\alpha}, S_{b\beta} \rangle = -\epsilon_{\alpha \beta} \delta_{ab}
= Str(S_{a\alpha} S_{b\beta}) 
\ee
It will turn out to be useful to choose the following basis of bosonic generators of $psu(1,1|2)$:
\beq K_{0} = \left( \begin{array}{cc} -\frac{i}{2}\sigma_3 &0 \\ 0&0 \end{array} \right) & 
K_{1} = \left( \begin{array}{cc} \frac{1}{2}\sigma_2 &0 \\ 0&0 \end{array} \right) &
K_{2} = \left( \begin{array}{cc} -\frac{1}{2}\sigma_1 &0 \\ 0&0 \end{array} \right) \\
K_{3} = \left( \begin{array}{cc} 0&0 \\ 0&-\frac{i}{2}\sigma_3 \end{array} \right) & 
K_{4} = \left( \begin{array}{cc} 0&0 \\ 0&-\frac{i}{2}\sigma_1 \end{array} \right) &
K_{5} = \left( \begin{array}{cc} 0&0 \\ 0&-\frac{i}{2}\sigma_2 \end{array} \right)
\eeq
These are related to the $K_{ab}$ by a simple linear change of basis:
\begin{align}
K_0 = \frac{1}{2}(K_{12}+K_{34}) \qquad K_1 = \frac{1}{2i}(K_{13}-K_{24}) \qquad K_2 = \frac{1}{2i}(K_{14}+K_{23}) \cr
K_3 = \frac{1}{2}(K_{12}-K_{34}) \qquad K_4 = \frac{1}{2}(K_{23}-K_{14}) \qquad K_5 = -\frac{1}{2}(K_{13}+K_{24}) \,.
\end{align}
These generators satisfy
\be 
\langle K_i, K_j \rangle = \frac{1}{2} \eta_{ij} =  Str(K_i K_j) 
\ee
with signature $(-+++++)$. For the fermionic generators, we use the linear combinations
\be
S^{\pm}_{1\a}= S_{1\a}\pm i S_{2\a} \qquad S^{\pm}_{3\a} = S_3 \pm i S_4 \,.
\ee
One can easily obtain the commutation relations between these generators from the ones in \eqref{KKcomm}. These will prove to be useful in writing out the action explicitly.  

\subsection{Explicit Form of the Worldsheet Currents} \label{componentaction}

Our goal in this section will be to obtain a concrete parameterization of the sigma-model on the
supergroup. We parameterize the group element $g \in SU(1,1|2)$ as \be
g = e^{\alpha} g_F g_{S^3} g_{AdS_3} \ee with \be g_F =
e^{\theta^{a\alpha}S_{a\alpha}} \ee
\beq g_{S^3} &=& e^{-(\varphi_1 + \varphi_2)K_3} e^{-2\theta K_4} e^{-(\varphi_1 - \varphi_2)K_3} \nonumber \\
&=& \left( \begin{array}{cc} \mathbb{I}&0\\ 0 & \begin{array}{cc} e^{i
        \varphi_1} \cos \theta & i e^{i\varphi_2}\sin \theta \\ i
      e^{-i \varphi_2}\sin \theta & e^{-i\varphi_1}\cos
      \theta \end{array} \end{array} \right) \eeq
and $\alpha$ is an overall phase that we will gauge away
eventually. For $AdS_3$ we choose the Poincar\'e parameterization: \beq g_{AdS_3}
&=& \frac{1}{2} \left( \begin{array}{cc}
\begin{array}{cc} e^{-\phi} + (\gamma-i) (\bar \gamma+i) e^\phi 
&  e^{-\phi}+ e^{\phi}(\gamma-i) (\bar \gamma-i) 
\\  
 e^{-\phi}+ e^{\phi}(\gamma+i) (\bar \gamma+i)  
& e^{-\phi} + (\gamma+i) (\bar \gamma-i) e^\phi 
\end{array} &0 \\ 0&\mathbb{I} 
\end{array} \right). \eeq
This follows from the following parameterization of $SL(2,\mathbb{R})$:
\beq g_{SL(2,\mathbb{R})} &=& 
\left( 
\begin{array}{cc} 
\begin{array}{cc} e^{-\phi} + \gamma \bar \gamma e^\phi &  e^{\phi} \gamma \\  e^{\phi} \bar \gamma & e^{\phi} 
\end{array} &0 \\ 0&\mathbb{I} 
\end{array} \right). \eeq
For Lorentzian $AdS_3$ both $\gamma$ and $\bar \gamma$ are real while they
become complex conjugate in Euclidean $AdS_3$.
We have used the following isomorphism from $SL(2,\mathbb{R})$ to $SU(1,1)$: 
\be
g \rightarrow cgc^\dagger ,
\ee
where the matrix $c$ is given by: 
\begin{eqnarray}
c &=& \frac{1}{\sqrt{2}} 
\left(
\begin{array}{cc}
1 & -i \\
1 & i 
\end{array}
\right).
\end{eqnarray}
We now expand the left-invariant one-form $g^{-1} dg$ on our basis of generators. We have:
\be g^{-1} dg = (g_{S^3} g_{AdS_3})^{-1} e^{-\theta^{a\alpha}F_{a\alpha}}d(e^{\theta^{a\alpha}F_{a\alpha}}) g_{S^3} g_{AdS_3} + g_{S^3}^{-1} d g_{S^3} + g_{AdS_3}^{-1} d g_{AdS_3} + d\alpha \mathbb{I} \ee
The $S^3$ part is given by 
\begin{align}\label{s3boscurrents} 
g_{S^3}^{-1} d g_{S^3} &= 2[\sin^2 \theta d \varphi_2 - \cos^2 \theta d \varphi_1]K_3 \cr
& + 2[-\cos(\varphi_1-\varphi_2) d\theta - \cos \theta \sin \theta \sin (\varphi_1-\varphi_2) (d\varphi_1 +d\varphi_2)] K_4 \cr
& + 2[-\sin(\varphi_1-\varphi_2) d\theta + \cos \theta \sin \theta \cos (\varphi_1-\varphi_2) (d\varphi_1 +d\varphi_2)] K_5 \cr
&= \sum_{i=3,4,5}j^i K_i \,.
\end{align}
One can check that:
\be \langle g_{S^3}^{-1} \p g_{S^3} , g_{S^3}^{-1} \bar \p g_{S^3} \rangle = 2 (\p \theta \bar \p \theta + \cos^2 \theta \p \varphi_1 \bar \p \varphi_1 + \sin^2 \theta \p \varphi_2 \bar \p \varphi_2) \,.\ee
So we recover the worldsheet action for a sigma-model with target space $S^3$. The other purely bosonic part is given by
\begin{align} \label{spherecurrents}
g_{AdS_3}^{-1} d g_{AdS_3} &= [d \bar \gamma + 2 \bar \gamma d \phi - e^{2\phi}(1+\bar \gamma^2)d \gamma]K_0 \cr
& + 2[d \phi-e^{2\phi}\bar \gamma d \gamma] K_1 \cr
& + [-d \bar \gamma-2\bar \gamma d \phi + e^{2\phi}(-1+\bar \gamma^2) d \gamma] K_2 \cr
&= \sum_{i=0,1,2}j^i K_i \,.
\end{align}
The action for the $AdS$ part thus takes the form
\be\label{ads3part} 
\langle g_{AdS_3}^{-1} \p g_{AdS_3} , g_{AdS_3}^{-1} \bar \p g_{AdS_3} \rangle = e^{2\phi} \left[ \bar \partial\bar \gamma \partial\gamma + \bar\partial \gamma \partial\bar \gamma\right] + 2\partial\phi \bar\partial\phi \,. 
\ee
The remaining term can be tackled as follows. First, let us consider
\begin{multline}
g_{F}^{-1}dg_{F} = d\theta^{a\alpha}S_{a\alpha} + \frac{1}{2}d\theta^{a\alpha}\theta^{b\beta} \epsilon_{\alpha\beta}\epsilon_{abcd}K_{cd} +  \frac{1}{2}d\theta^{a\alpha}\theta^{b\beta}\theta^{c\gamma} \epsilon_{\alpha\beta}\epsilon_{abcd}S_{d\gamma} \cr
+\frac{1}{6}d\theta^{a\alpha}\theta^{b\beta}\theta^{c\gamma}\epsilon_{\alpha\beta}
\epsilon_{\gamma\delta}\big[K_{ac} \theta^{b\delta}+K_{cb}\theta^{a\delta}+K_{ba}\theta^{c\delta}\big] + \ldots \,. 
\end{multline}
Let us write this compactly as follows:
\be\label{deffandb}
g_{F}^{-1}dg_{F} = f^{a\alpha} S_{a\alpha} + b^{i} K_{i}
\ee
In order to obtain the left-invariant one-forms, it is necessary
to obtain the result of conjugating the generators by the bosonic
group elements. An economical way to express the conjugation relations
is as follows:
\begin{align}
e^{x K_{ab}}S_{c\alpha}e^{-xK_{ab}} &= (\cos x S_{a\alpha} +  \sin x S_{b\alpha})\quad \text{for}\quad a=c \quad\text{and}\quad b\ne c \cr
e^{x K_{ab}}K_{cd}e^{-xK_{ab}} &= 
(\cos x K_{ad} + \sin x K_{bd}) \quad \text{for}\quad a=c \quad\text{and}\quad b\ne d \,.
\end{align}
Recalling that $K_{ab}=-K_{ba}$, we can obtain all other possibilities easily. This proves sufficient to compute the bosonic and fermionic currents. 

\subsubsection*{Fermionic Currents}

Let us first compute the fermionic currents 
\begin{align} 
\Pi_{a\alpha} &=   i\, \langle g^{-1} dg, S_{a\alpha} \rangle \,. 
\end{align}
For this we have to compute 
\be
g_{B}^{-1} S^{\pm}_{a\alpha}g_{B} =  (g_{AdS})^{-1}\left[ (g_{S^3})^{-1} S^{\pm}_{a\a}\ g_{S^3} \right] g_{AdS} \,.
\ee
Then,
\be
\Pi_{a\a} = \epsilon_{\a\b} f^{b\b}\delta_{ab} \,.
\ee
In this calculation, we find it convenient to use the
global $(t,\varphi,\rho)$ parameterization of $AdS_3$:
\be
g_{AdS_3} = e^{-(t+\varphi)K_0}\, e^{-2\rho K_2}\, e^{-(t-\varphi)K_0} \,.
\ee
Under conjugation by the bosonic group element, the transformation of the fermionic generators is encoded in a matrix ${\cal M}$ such that 
\be
\begin{pmatrix} S^{+}_{1\alpha} \cr S^{-}_{1\alpha} \cr S^{+}_{3\alpha} \cr S^{-}_{3\alpha} \end{pmatrix} \longrightarrow {\cal M} \begin{pmatrix} S^{+}_{1\alpha} \cr S^{-}_{1\alpha} \cr S^{+}_{3\alpha} \cr S^{-}_{3\alpha} \end{pmatrix} \,.
\ee
where the matrix ${\cal M}$ is given by
\be
\begin{pmatrix}
e^{-i(t+\varphi_1)}c_{\theta}\cosh\rho & -i e^{-i(\varphi + \varphi_2)}s_{\theta}\sinh\rho & i e^{-i(t+\varphi_2)}\cosh\rho s_{\theta} & e^{-i(\varphi+\varphi_1)}c_{\theta}\sinh\rho \cr
-i e^{i(\varphi + \varphi_2)}s_{\theta}\sinh\rho & e^{i(t+\varphi_1)}c_{\theta}\cosh\rho & -e^{i(\varphi+\varphi_1)}c_{\theta}\sinh\rho & -i e^{i(t+\varphi_2)}\cosh\rho s_{\theta} \cr
i e^{-i(t-\varphi_2)}\cosh\rho s_{\theta} & -e^{-i(\varphi-\varphi_1)}c_{\theta}\sinh\rho & e^{-i(t-\varphi_1)}c_{\theta}\cosh\rho & i e^{-i(\varphi-\varphi_2)}s_{\theta}\sinh\rho \cr
e^{i(\varphi-\varphi_1)}c_{\theta}\sinh\rho &  -i e^{i(t-\varphi_2)}\cosh\rho s_{\theta} & i e^{i(\varphi-\varphi_2)}s_{\theta}\sinh\rho &  e^{i(t-\varphi_1)}c_{\theta}\cosh\rho
\end{pmatrix} \,.
\ee
We have denoted $s_{\theta}=\sin\theta$ and $c_{\theta}=\cos\theta$.
 The fermionic currents can then be written in the form
\be
\Pi^{\pm}_{a\a} = f^{a\beta\pm} {\cal M}_{ab}\epsilon_{\b\a} \,.
\ee

\subsubsection*{Bosonic Currents}

It is much easier to directly work with the $K_i$ generators than with the $K_{ab}$ generators to derive the contributions to the bosonic currents from the fermionic ($\theta$-dependent) terms in the action. Starting from \eqref{deffandb} and using the fact that $\{K_{0,1,2}\}$ commutes with $\{K_{3,4,5}\}$, we obtain this bosonic contribution to be
\be
g_{AdS_3}^{-1} \left[\sum_{i=0,1,2}b^i K_i \right] g_{AdS_3} + g_{S_3}^{-1} \left[\sum_{i=3,4,5}b^i K_i \right] g_{S_3} = \sum_{i} \ell^i K_i \,.
\ee
Using the commutation relations, we find that 
\begin{align}\label{allfermcurrents}
  \ell^0 &= \cosh2\rho\ b_0 \sinh2\rho\cos(t+\phi)\ b_1 -
  \sin(t+\phi)\sinh2\rho\ b_2 \cr \ell^1 &= \sinh2\rho \cos(t-\phi)\
  b_0 + (\cosh2\rho\cos(t+\phi)\cos(t-\phi)-\sin(t+\phi)\sin(t-\phi))\
  b_1 \cr
  &\hspace{1in}-(\cos(t+\phi)\sin(t-\phi)+\sinh2\rho\sin(t+\phi)\cos(t-\phi))\
  b_2\cr \ell^2 &= \sinh2\rho\sin(t-\phi)\ b_0
  +(\sin(t+\phi)\cos(t-\phi)+\cosh2\rho\cos(t+\phi)\sin(t-\phi))\ b_1
  \cr &\hspace{1in}+
  (\cos(t+\phi)\cos(t-\phi)-\cosh2\rho\sin(t+\phi)\sin(t-\phi) )\
  b_2\cr \ell^3 &= \cos2\theta\
  b_3+\sin(\varphi_1+\varphi_2)\sin2\theta\ b_4+
  \cos(\varphi_1+\varphi_2)\sin2\theta\ b_5\cr 
  \ell^4 &=\sin2\theta\sin(\varphi_1-\varphi_2)\ b_3\cr
  &\hspace{1in}+(\cos(\varphi_1+\varphi_2)\cos(\varphi_1-\varphi_2)  -\sin(\varphi_1+\varphi_2)\cos2\theta \sin(\varphi_1-\varphi_2))\ b_4\cr
&\hspace{1.5in}  
 -(\sin(\varphi_1+\varphi_2)\cos(\varphi_1-\varphi_2) +\cos(\varphi_1+\varphi_2) \cos2\theta \sin(\varphi_1-\varphi_2))\ b_5\cr 
  \ell^5&= -\sin2\theta\cos(\varphi_1-\varphi_2)\ b_3 \cr
&\hspace{1in}+ (\cos(\varphi_1+\varphi_2)\sin(\varphi_1-\varphi_2)+\sin(\varphi_1\varphi_2)\cos2\theta\cos(\varphi_1+\varphi_2))\  b_4\cr
&\hspace{1.5in}+(\cos(\varphi_1+\varphi_2)\cos2\theta\cos(\varphi_1-\varphi_2)  - \sin(\varphi_1+\varphi_2)\sin(\varphi_1-\varphi_2))\ b_5 \,.\cr
\end{align}

\subsection{The Action}
Once the $f$'s and $b$'s are calculated, then, using the explicit
components of the matrix ${\cal M}$ to simplify the fermionic currents, the
kinetic part of the Lagrangian (up to the overall constants in
\eqref{ourmodel}) can be written in the form
\be\label{fulllagrange} {\cal L} = \sum_{m,n=0}^{5}\eta_{mn}(\ell+j)^m
(\ell+j)^n + \sum_{a,b,\a,\b}\epsilon_{\a\b}\delta_{ab}f^{a\a}f^{b\b}
\,.  \ee
All the currents have been explicitly written out above and, 
one can write out the full action to all orders in the
fermionic coordinates.

\subsubsection*{Normalization}
 In order to check the normalization, it is useful to focus on the purely bosonic part. Let
us first write the worldsheet action explicitly in terms of the
Poincar\'e coordinates $(\gamma, \bar\gamma, \phi)$. Taking into account
$\eta^{z\zbar}=2$ and rewriting the partial derivatives of $g^{-1}$ in
terms of that on $g$, we get
 \be
 S_{WS} = \frac{1}{4\pi f^2}\int d^2z\, \text{STr}(g^{-1}\p g\, g^{-1}\pbar g) + S_{WZW}\,. 
 \ee
 Using equation \eqref{ads3part} and combining this with an antisymmetric part coming from the WZW term, we find that
 \begin{align}
 S_{WS} &= \frac{1}{4\pi}\int d^2z \left\{e^{2\phi}\left[\left(\frac{1}{f^2}+k\right)\p\bar\gamma\pbar\gamma + \left(\frac{1}{f^2}-k\right) \p\gamma\pbar\bar\gamma\right] + \frac{2}{f^2}\p\phi\pbar\phi \right\} +\ldots \cr
 &= \frac{1}{2\pi}\int d^2z 
 \left(e^{2\phi}
 \left[-c_+\p\bar\gamma\pbar\gamma -c_- \p\gamma\pbar\bar\gamma \right] + \frac{1}{f^2}\p\phi\pbar\phi 
 \right) +\ldots \,.
 \end{align}
 One can check that, at the WZW point, putting $f^{2}=1/k$, the action
 coincides with the bosonic action on $AdS_3$ with NS-NS flux
 with an overall multiplicative factor of $k/2\pi$.

\section{Primary Operators}\label{primaries}

We define a primary field $\phi$ with respect to the left-current algebra \eqref{euclidOPEs} as a field satisfying the OPEs:
\be j^a_{L,z}(z,\bar z) \phi(w,\bar w) = -\frac{c_+}{c_++c_-} \frac{t^a \phi(w,\bar w)}{(z-w)} + \text{less singular} \ee
\be j^a_{L,\bar z}(z,\bar z) \phi(w,\bar w) = -\frac{c_-}{c_++c_-} \frac{t^a \phi(w,\bar w)}{(z-w)} + \text{less singular} \ee
where $t^a$ is a generator of the Lie super-algebra in the
representation in which $\phi$ transforms on the left. The structure
of these OPEs is postulated, and the exact values of the coefficients
is fixed by demanding compatibility both with current conservation and
with the Maurer-Cartan equation.  We will now show that this
definition implies that a primary field with respect to the
left-current algebra is also a primary field with respect to the
Virasoro algebra.  The worldsheet stress tensor is\footnote{We will
  not be careful about minus signs due to fermionic statistics. They
  can be consistently restored.}: \be T(z) =
\frac{1}{2c_1}\kappa_{ab}:j^a_{L,z} j^b_{L,z}:(z). \ee Let us compute
the OPE between a left-primary field $\phi$ and the holomorphic
stress-tensor:
\begin{align} 
\phi(z) T(w) &= \frac{1}{2 c_1}\lim_{:x \to w:}\phi(z) j_{aL,z}(x) j^a_{L,z}(w) \cr
&= \frac{1}{2c_1}\lim_{:x \to w:} \left(
-\frac{c_+}{c_++c_-} \frac{t_a \phi(x)}{x-z}j^a_{L,z}(w) -  \frac{c_+}{c_++c_-} j^a_{L,z}(x)\frac{t_a \phi(w)}{w-z} 
\right)\cr
&=  \frac{c_+}{2c_1(c_++c_-)}\lim_{:x \to w:} \left(
 \frac{t_a}{x-z} \left(\frac{c_+}{c_++c_-} \frac{t^a \phi(w)}{w-x}- : \phi j^a_{L,z}:(w) \right) \right. \cr
& \left. \hspace{1.5 in}+ \frac{t_a}{w-z} \left(\frac{c_+}{c_++c_-} \frac{t^a \phi(w)}{x-w}- : j^a_{L,z} \phi :(w) \right)
\right) \cr
&=  \frac{c_+}{2c_1(c_++c_-)} \left(
\frac{c_+}{c_++c_-}\frac{t_a t^a \phi(w)}{(z-w)^2} - 2 \frac{t_a : j^a_{L,z} \phi :(w)}{w-z} - \frac{c_+}{c_++c_-} \frac{t_a t^a \partial \phi(w)}{w-z} \right) \cr
& = \frac{f^2}{2}\frac{t_a t^a \phi(w)}{(z-w)^2} + \frac{1}{c_+} \frac{t_a : j^a_{L,z} \phi :(w)}{w-z} -\frac{f^2}{2}\frac{t_a t^a \p \phi(w)}{w-z}.
\end{align}
We deduce the OPE between the stress-tensor and the primary field $\phi$:
\be T(w) \phi(z) = \frac{f^2}{2} \frac{t_a t^a \phi(z)}{(w-z)^2}
 + \frac{1}{c_+}  \frac{t_a : j^a_{L,z} \phi :(z)}{w-z}. \ee
 We observe that there is no pole of order greater than two, and that
 the pole of order two is proportional to the operator $\phi$. This
 implies that the operator $\phi$ is annihilated by all the positive
 modes of the stress tensor, and thus that it is a Virasoro primary.
 We can read off the conformal dimension of the operator $\phi$:
\be \Delta_{\phi}  = \frac{f^2}{2} t_a t^a. \ee
The simple pole in the $T.\phi$ OPE gives the holomorphic derivative of $\phi$:
\be\label{dPrimary} \p \phi = \frac{1}{c_+}  t_a : j^a_{L,z} \phi :(z). \ee

\section{The Virasoro Algebra}\label{TTOPE}

In this appendix we will consider the self-OPE of the stress tensor \eqref{TTOPEeqn}. As we did in the bulk of the paper, we will evaluate the OPE of the stress-tensor with its anti-holomorphic derivative :
\begin{multline}\label{TTstep1}
\p_{\xbar}T(x)\cdot T(y) = -\frac{1}{4\pi i}\int d^2 w\cr 
\times \left[\oint dz (\p_xj_{L,z}\p_x\Phi_1 + 2\p_x^2j_{L,z}\Phi_1) + \oint d\zbar (\p_xj_{L,\zbar}\p_x\Phi_1 + 2\p_x^2j_{L,\zbar}\Phi_1) \right] \cr
\times\left[ (\p_yj_{L,z}\p_y\p_{\bar w} \Lambda + 2\p_y^2j_{L,z}\p_{\bar w} \Lambda) + (\p_yj_{L,\zbar}\p_y\p_{\bar w} \Lambda + 2\p_y^2j_{L,\zbar}\p_{\bar w} \Lambda)\right]
\end{multline}
where we kept the coordinate dependence of the operators implicit. In
order to compute this OPE, we require the OPEs between the
$x$-dependent combinations of the currents (and their derivatives),
which we list below for convenience:
\begin{align}
\p_xj_{L,z}\cdot \p_y j_{L,z} &\sim \frac{-2c_1}{(z-w)^2} + \frac{2c_2(x-y)\p_y^2j_{L,z}}{i(z-w)} + \frac{2(c_2-g)(\zbar-\wbar)(x-y)\p_y^2j_{L,\zbar}}{i(z-w)^2} \cr
\p_xj_{L,z}\cdot \p_y^2j_{L,z} &\sim \frac{-2c_2\p_y^2j_{L,z}}{i(z-w)} + \frac{-2(c_2-g)(\zbar-\wbar)\p_y^2j_{L,\zbar}}{i(z-w)^2} \cr
\p_xj_{L,z}\cdot \p_yj_{L,\zbar} &\sim -2\tilde{c}\delta^{(2)}(z-w) + \frac{2(c_4-g)f^{ab}_c(x-y)\p_y^2j_{L,z}}{i(\zbar-\wbar)} + \frac{2(c_2-g)f^{ab}_c(x-y)\p_y^2j_{L,\zbar}}{i(z-w)} \cr
\p_xj_{L,z}\cdot \p_y^2j_{L,\zbar} &\sim \frac{-2(c_4-g)f^{ab}_c\p_y^2j_{L,z}}{i(\zbar-\wbar)} + \frac{-2(c_2-g)f^{ab}_c\p_y^2j_{L,\zbar}}{i(z-w)} \cr
\p_x^2j_{L,z}\cdot \p_yj_{L,z} &\sim \frac{2c_2\p_y^2j_{L,z}}{i(z-w)} + \frac{2(c_2-g)(\zbar-\wbar)\p_y^2j_{L,\zbar}}{i(z-w)^2} \cr
\p_x^2j_{L,z}\cdot \p_yj_{L,\zbar} &\sim \frac{2(c_4-g)f^{ab}_c\p_y^2j_{L,z}}{i(\zbar-\wbar)} + \frac{2(c_2-g)f^{ab}_c\p_y^2j_{L,\zbar}}{i(z-w)} \cr
\p_xj_{L,z} \Phi_1 &\sim \frac{c_+}{c_+ +c_-} \frac{1}{z-w}[2(x-y)\p_y -2]\Phi_1 \cr
\p_xj_{L,z} \p_y\Phi_1 &\sim \frac{c_+}{c_+ +c_-} \frac{1}{z-w}[2(x-y)\p_y^2 -4\p_y]\Phi_1 \cr
\p_x^2j_{L,z} \Phi_1 &\sim \frac{c_+}{c_+ +c_-} \frac{2\p_y\Phi_1}{z-w} \cr
\p_x^2j_{L,z} \p_y\Phi_1 &\sim \frac{c_+}{c_+ +c_-} \frac{2\p_y^2\Phi_1}{z-w} \,.
\end{align}
The OPEs of the current component $j_{L,\zbar}$ with the operator
$\Phi_1$ are identical except the overall factor outside is
$c_-/(c_++c_-)$ and the antiholomorphic factor $1/(\zbar-\wbar)$
replaces the holomorphic factor.  The contour integrals in equation
\eqref{TTstep1} will pick up the poles in the OPEs between the
integrated operators.  One useful point to note is that the mixed
terms in the OPEs, which contain both holomorphic and anti-holomorphic
pieces in $z-w$ do not contribute to the contour integrals.  The full
computation is quite tedious.  The key point is that because of the
relations \eqref{relationsCs} satisfied by the coefficients of the
current algebra, the computation is a simple non-chiral
generalization of the calculation for the case with pure NS flux
\cite{Kutasov:1999xu}.

Let us illustrate this for the most singular term in the above OPE.
It is obtained by taking the most singular terms in the
current-current OPEs (i.e. the doubles poles and the contact
terms). Collecting the four relevant terms in equation \eqref{TTstep1}, we get
:
\begin{multline}
-\frac{1}{4\pi i}\int d^2w \oint dz\left[ \frac{-2c_1}{(z-w)^2} \p_x\Phi_1\cdot \p_y \p_{\bar w}\Lambda -2 \tilde{c}\delta^{(2)}(z-w)\p_x\Phi_1\cdot \p_y \p_w \Lambda\right] \cr
-\frac{1}{4\pi i}\int d^2 w \oint d\zbar\left[\frac{-2c_3}{(\zbar-\wbar)^2} \p_x\Phi_1\cdot \p_y \p_w \Lambda -2 \tilde{c}\delta^{(2)}(z-w)\p_x\Phi_1\cdot \p_y \p_{\bar w} \Lambda \right] \,. 
\end{multline}
Let us first combine the first and the fourth terms, followed by the
second and the third terms. After performing the contour integration,
this leads to
\begin{align}
 \int d^2w &\left[ \frac{(c_1-\tilde{c})}{c_+} \lim_{:z \to w:}\left[\p_x^2(j_{L,z} \Phi_1)(x;w)\cdot \p_y \p_{\bar w} \Lambda(y;w)\right] 
\right. \cr
&\left. +  \frac{(c_3-\tilde{c})}{c_-} \lim_{:z \to w:}\left[\p_x^2(j_{L,\zbar} \Phi_1)(x;w)\cdot \p_y \p_w \Lambda(y;w)\right] \right]
\end{align}
The coefficients simplify thanks to the relations \eqref{relationsCs}.
The regular limit leads to a $\delta^{(2)}(x-y)$ factor and the most
singular term in the OPE \eqref{TTstep1} occurs when both partial
derivatives $\p_x^2$ act on it. Finally we obtain :
\be
\p_{\xbar}T(x)\cdot T(y) =- \pi \p_x^3\delta^{(2)}(x-y) I + \ldots \,,
\ee
where $I$ is the operator defined in equation \eqref{1stDefI} that we already encountered in the self-OPE of the R-current.

Integrating with respect to $\xbar$, we find 
\be
T(x)\cdot T(y) = \frac{3I}{(x-y)^4} + \ldots  
\ee
which gives $c = 6 \, I$. This generalizes the result of
\cite{Kutasov:1999xu} to $AdS_3 \times S^3$ backgrounds which include
RR fluxes. A similar analysis can be done for all the other
terms leading to the standard operator product expansion for 
the energy-momentum tensor.

\end{appendix}


\begin{thebibliography}{99}
\bibitem{Maldacena:1997re}
  J.~M.~Maldacena,
  ``The large N limit of superconformal field theories and supergravity,''
  Adv.\ Theor.\ Math.\ Phys.\  {\bf 2} (1998) 231
  [Int.\ J.\ Theor.\ Phys.\  {\bf 38} (1999) 1113]
  [arXiv:hep-th/9711200].

\bibitem{Berenstein:2002jq}
  D.~E.~Berenstein, J.~M.~Maldacena and H.~S.~Nastase,
  ``Strings in flat space and pp waves from N = 4 super Yang Mills,''
  JHEP {\bf 0204} (2002) 013
  [arXiv:hep-th/0202021].

\bibitem{Metsaev:2002re}
  R.~R.~Metsaev and A.~A.~Tseytlin,
  ``Exactly solvable model of superstring in plane wave Ramond-Ramond
  background,''
  Phys.\ Rev.\  D {\bf 65} (2002) 126004
  [arXiv:hep-th/0202109].

\bibitem{Minahan:2002ve}
  J.~A.~Minahan and K.~Zarembo,
  ``The Bethe-ansatz for N = 4 super Yang-Mills,''
  JHEP {\bf 0303} (2003) 013
  [arXiv:hep-th/0212208].

\bibitem{Bena:2003wd}
  I.~Bena, J.~Polchinski and R.~Roiban,
  ``Hidden symmetries of the AdS(5) x S**5 superstring,''
  Phys.\ Rev.\  D {\bf 69}, 046002 (2004)
  [arXiv:hep-th/0305116].

\bibitem{Gromov:2009tv}
  N.~Gromov, V.~Kazakov and P.~Vieira,
  ``Integrability for the Full Spectrum of Planar AdS/CFT,''
  arXiv:0901.3753 [hep-th].

\bibitem{Brown:1986nw}
  J.~D.~Brown and M.~Henneaux,
  ``Central Charges in the Canonical Realization of Asymptotic Symmetries: An
  Example from Three-Dimensional Gravity,''
  Commun.\ Math.\ Phys.\  {\bf 104}, 207 (1986).

\bibitem{Giveon:1998ns}
  A.~Giveon, D.~Kutasov and N.~Seiberg,
  ``Comments on string theory on AdS(3),''
  Adv.\ Theor.\ Math.\ Phys.\  {\bf 2} (1998) 733
  [arXiv:hep-th/9806194].



\bibitem{de Boer:1998pp}
  J.~de Boer, H.~Ooguri, H.~Robins and J.~Tannenhauser,
  ``String theory on AdS(3),''
  JHEP {\bf 9812} (1998) 026
  [arXiv:hep-th/9812046].

\bibitem{Kutasov:1999xu}
  D.~Kutasov and N.~Seiberg,
  ``More comments on string theory on AdS(3),''
  JHEP {\bf 9904} (1999) 008
  [arXiv:hep-th/9903219].

\bibitem{BVW}
  N.~Berkovits, C.~Vafa and E.~Witten,
  ``Conformal field theory of AdS background with Ramond-Ramond flux,''
  JHEP {\bf 9903}, 018 (1999)
  [arXiv:hep-th/9902098].

\bibitem{ABT}
  S.~K.~Ashok, R.~Benichou and J.~Troost,
  ``Conformal Current Algebra in Two Dimensions,''
  arXiv:0903.4277 [hep-th].

\bibitem{Giveon:2001up}
  A.~Giveon and D.~Kutasov,
  ``Notes on AdS(3),''
  Nucl.\ Phys.\  B {\bf 621}, 303 (2002)
  [arXiv:hep-th/0106004].

\bibitem{Bershadsky:1999hk}
  M.~Bershadsky, S.~Zhukov and A.~Vaintrob,
  ``PSL(n|n) sigma model as a conformal field theory,''
  Nucl.\ Phys.\  B {\bf 559} (1999) 205
  [arXiv:hep-th/9902180].

\bibitem{Gotz:2005ka}
  G.~Gotz, T.~Quella and V.~Schomerus,
  ``Tensor products of psl(2 2) representations,''
  arXiv:hep-th/0506072.

\bibitem{Gotz:2006qp}
  G.~Gotz, T.~Quella and V.~Schomerus,
  ``The WZNW model on PSU(1,1 2),''
  JHEP {\bf 0703}, 003 (2007)
  [arXiv:hep-th/0610070].

\bibitem{Quella:2007sg}
  T.~Quella, V.~Schomerus and T.~Creutzig,
  ``Boundary Spectra in Superspace Sigma-Models,''
  JHEP {\bf 0810} (2008) 024
  [arXiv:0712.3549 [hep-th]].

\bibitem{Berkovits:1994vy}
  N.~Berkovits and C.~Vafa,
  ``N=4 topological strings,''
  Nucl.\ Phys.\  B {\bf 433} (1995) 123
  [arXiv:hep-th/9407190].

\bibitem{Berkovits:1999du}
  N.~Berkovits,
  ``Quantization of the type II superstring in a curved six-dimensional background,''
  Nucl.\ Phys.\  B {\bf 565}, 333 (2000)
  [arXiv:hep-th/9908041].


\bibitem{Dolan:1999dc}
  L.~Dolan and E.~Witten,
  ``Vertex operators for AdS(3) background with Ramond-Ramond flux,''
  JHEP {\bf 9911}, 003 (1999)
  [arXiv:hep-th/9910205].

\bibitem{de Boer:1998ip}
  J.~de Boer,
  ``Six-dimensional supergravity on S**3 x AdS(3) and 2d conformal field
  theory,''
  Nucl.\ Phys.\  B {\bf 548} (1999) 139
  [arXiv:hep-th/9806104].

\bibitem{Henneaux:1999ib}
  M.~Henneaux, L.~Maoz and A.~Schwimmer,
  ``Asymptotic dynamics and asymptotic symmetries of three-dimensional
  extended AdS supergravity,''
  Annals Phys.\  {\bf 282} (2000) 31
  [arXiv:hep-th/9910013].

\bibitem{Teschner:1999ug}
  J.~Teschner,
  ``Operator product expansion and factorization in the H-3+ WZNW model,''
  Nucl.\ Phys.\  B {\bf 571} (2000) 555
  [arXiv:hep-th/9906215].

\bibitem{Aharony:2007rq}
  O.~Aharony and Z.~Komargodski,
  ``The space-time operator product expansion in string theory duals of field theories,''
  JHEP {\bf 0801} (2008) 064
  [arXiv:0711.1174 [hep-th]].


\bibitem{Strominger:1997eq}
  A.~Strominger,
  ``Black hole entropy from near-horizon microstates,''
  JHEP {\bf 9802} (1998) 009
  [arXiv:hep-th/9712251].




\bibitem{yellowbook}
P.~Di Francesco, P.~Mathieu, D.~Senechal, ``Conformal Field Theory," Springer, 1997.




\bibitem{Henningson:1998gx}
  M.~Henningson and K.~Skenderis,
  ``The holographic Weyl anomaly,''
  JHEP {\bf 9807} (1998) 023
  [arXiv:hep-th/9806087].

\bibitem{Berkovits:1994wr}
  N.~Berkovits,
  ``Covariant quantization of the Green-Schwarz superstring in a Calabi-Yau background,''
  Nucl.\ Phys.\  B {\bf 431} (1994) 258
  [arXiv:hep-th/9404162].


\bibitem{Berkovits:1999zq}
  N.~Berkovits, M.~Bershadsky, T.~Hauer, S.~Zhukov and B.~Zwiebach,
  ``Superstring theory on AdS(2) x S(2) as a coset supermanifold,''
  Nucl.\ Phys.\  B {\bf 567} (2000) 61
  [arXiv:hep-th/9907200].

\end{thebibliography}
\end{document}